\documentclass[aps,prd,nofootinbib,preprint,floatfix,unsortedaddress]{revtex4-1}
\usepackage{setspace}
\usepackage{graphicx}
\usepackage{dcolumn}
\usepackage{multirow}
\usepackage{subfigure}
\usepackage{times,mathptm}
\usepackage{float}
\usepackage{color}
\usepackage{amsmath,amsfonts}
\usepackage{mathptmx}
\usepackage{mathrsfs}
\usepackage{bbm}
\usepackage{bm}
\usepackage{xfrac}

\newcommand{\beq}{\begin{equation}}
\newcommand{\eeq}{\end{equation}} 
\newcommand{\bea}{\begin{eqnarray}}
\newcommand{\eea}{\end{eqnarray}}

\renewcommand{\d}{\delta}

\renewcommand{\l}{\lambda}

\newcommand{\tQ}{\widetilde{Q}}
\newcommand{\tK}{\widetilde{K}}

\renewcommand{\b}{\beta}
\renewcommand{\a}{\alpha}

\newcommand{\tr}{\text{Tr}}

\newcommand{\bx}{\mathbf{x}}

\newcommand{\vx}{{\vec{x}}}
\newcommand{\vy}{{\vec{y}}}

\newcommand{\vk}{{\vec{k}}} 
\newcommand{\vq}{{\vec{q}}}

\newcommand{\m}{\mu}

\newcommand{\s}{\sigma}
\renewcommand{\k}{\kappa}

\newcommand{\D}{\Delta}

\renewcommand{\th}{\theta}

\newcommand{\oh}{\frac{1}{2}}
\newcommand{\oq}{\frac{1}{4}}
\newcommand{\ot}{\frac{1}{3}}
\newcommand{\on}{\frac{1}{9}}
\newcommand{\dg}{\dagger}
\newcommand{\non}{\nonumber}

\newcommand{\rf}[1]{(\ref{#1})}
\newcommand{\ra}{\rightarrow}
\newcommand{\pa}{\partial}
\renewcommand{\vec}[1]{\bm #1}

\usepackage{ulem}

\begin{document}

\title{Finding the effective Polyakov line action for \\ SU(3) gauge theories at finite chemical potential} 
 
\author{Jeff Greensite}
\affiliation{\singlespacing Physics and Astronomy Department, \\ San Francisco State
University, San Francisco, CA~94132, USA}

\author{Kurt Langfeld}
\affiliation{\singlespacing School of Computing \& Mathematics, University of Plymouth, Plymouth, PL4 8AA, UK}
\date{\today}
\vspace{60pt}
\begin{abstract}

\singlespacing
 
      Motivated by the sign problem, we calculate the effective Polyakov line action corresponding to certain SU(3) lattice gauge theories on a ${16^3 \times 6}$ lattice via the ``relative weights'' method introduced in our previous articles.   The calculation
is carried out  at $\b=5.6,5.7$ for the pure gauge theory, and at $\b=5.6$ for the gauge field coupled to a relatively light scalar particle.   In the latter example we determine the effective theory also at finite chemical potential, and show how observables 
relevant to phase structure can be computed in the effective theory via mean field methods.  In all cases a comparison of
Polyakov line correlators in the effective theory and the underlying lattice gauge theory, computed numerically at zero chemical potential, shows accurate agreement down to correlator magnitudes of order $10^{-5}$.  We also derive the effective Polyakov line action corresponding to a gauge theory with heavy quarks and large chemical potential, and apply mean field methods to extract observables.

\end{abstract}

\maketitle

\singlespacing
\section{\label{intro}Introduction}

The effective Polyakov line action $S_P$ of a lattice gauge theory is defined by integrating out all degrees of freedom
of the lattice gauge theory, under the constraint that the Polyakov line holonomies are held fixed.  It is convenient to implement this constraint in temporal gauge (${U_0(\bx,t\ne 0)=\mathbbm{1}}$), so that
\bea
\exp\Bigl[S_P[U_{\vx},U^\dg_{\vx}]\Bigl] =    \int  DU_0(\vx,0) DU_k  D\phi ~ \left\{\prod_{\vx} \d[U_{\vx}-U_0(\vx,0)]  \right\}
 e^{S_L} \ ,
\label{S_P}
\eea
where $\phi$ denotes any matter fields, scalar or fermionic, coupled to the gauge field, and $S_L$ is the lattice action (note that we adopt a sign convention for the Euclidean action such that the Boltzman weight is proportional to 
$\exp[+S]$).\footnote{Temporal gauge is convenient but not essential. In the absence of gauge-fixing one could simply apply the Metropolis algorithm to simultaneous trial updates of neighboring timelike links
 $U_0(\vx,t)\ra G U_0(\vx,t), U_0(\vx,t-1) \ra U_0(\vx,t-1)G^\dg$ , where $G$ is an SU(3) group element, which leave the Polyakov line holonomy fixed.}   The effective Polyakov line action 
$S_P$ can be computed analytically from the underlying lattice gauge theory at strong gauge couplings and heavy quark masses, and at leading order it has the form of an SU(3) spin model in $D=3$ dimensions\footnote{$S_P$ has been computed to higher orders in the combined strong-coupling/hopping parameter expansion in ref.\ \cite{Fromm:2011qi}.}
\bea
          S_{spin} =  J \sum_x \sum_{k=1}^{3}\Bigl(\tr[U_x] \tr[U^\dg_{x+\hat{k}}] + \text{c.c.}\Bigr) +
                                h \sum_x \Bigl( e^{\m/T} \tr[U_x] + e^{-\m/T} \tr[U^\dg_x] \Bigr) \ .
\label{spin}
\eea
This model has been solved at finite chemical potential $\m$ by several different methods, including the flux representation 
\cite{Mercado:2012ue}, stochastic
quantization \cite{Aarts:2011zn}, reweighting \cite{Fromm:2011qi}, and the mean field approach \cite{Greensite:2012xv}.   

    This article is concerned with computing $S_P$ from the underlying lattice gauge theory at gauge couplings which are not so strong, and matter fields which are not so heavy.  The motivation is that since the phase diagram for $S_{spin}$ has been determined over a large range of $J,h,\m$ by the methods mentioned above, perhaps the same methods can be successfully applied to solve $S_P$, providing that theory is known in the parameter range (of temperature, quark mass, and chemical potential) of interest.  The phase diagram of the effective theory will mirror the phase diagram of the underlying gauge theory.  
    
    There is a simple relationship between the effective Polyakov line  action (PLA) $S^{\m=0}_P$ corresponding to zero chemical potential in the underlying lattice gauge theory, and the PLA $S^\m_P$ corresponding to finite $\m$ in the underlying theory: 
\bea
     S_P^\m[U_\vx,U^\dg_\vx] =  S_P^{\m=0}[e^{N_t \m} U_\vx,e^{-N_t \m}U^\dg_\vx]   \ .
\label{convert}
\eea
This relationship was shown to be true to all orders in the strong-coupling/hopping parameter expansion  
\cite{Greensite:2013yd}; presumably it holds in general.   However, if $S_P$ is expressed in terms of the trace of
Polyakov line holonomies, rather than the holonomies themselves, then certain ambiguities arise in the use of \rf{convert}.  We will show how these ambiguities are resolved by computing the PLA numerically also at imaginary values $\m = i\th/N_t$ of the chemical potential. 

    In order to determine the PLA at $\m=0$ (and at imaginary $\m$) we make use of the ``relative weights'' method, 
which was introduced and tested on SU(2) lattice gauge theory in our two previous articles on this subject \cite{Greensite:2013yd,Greensite:2013bya}.  There is no sign problem for the SU(2) gauge group, but it is still a challenge to extract the PLA from the underlying gauge theory.  The criterion for success of the method, for any SU(N) gauge group, is that spin-spin correlators
\beq
             G(R) = {1\over N^2} \langle \tr[U_\vx] \tr[U_\vy^\dg] \rangle ~~,~~ R=|\vx-\vy|
\label{GR}
\eeq
computed in the effective theory agree with the corresponding Polyakov line correlators in the underlying lattice gauge theory.  For SU(2) lattice gauge theory we found agreement at gauge couplings ranging from very strong couplings up to the deconfinement transition, for separations $R$ up to twelve lattice spacings, and over a range of correlator values down to 
$O(10^{-5})$.\footnote{There have been other approaches to the problem of determining the PLA, notably the Inverse Monte Carlo method \cite{Dittmann:2003qt,*Heinzl:2005xv}, and strong-coupling expansions \cite{Fromm:2011qi}, but these have so far not demonstrated an agreement in the Polyakov line correlators beyond separations of two or three lattice spacings
(for recent work, see \cite{Bergner:2013rea}).}  In this article we will extend our previous work to the SU(3) gauge theory.  It is ultimately our intention to compute the effective PLA for gauge fields coupled to light quarks.  However, in this first investigation, we prefer to avoid the complexities of dynamical fermion simulations and study instead the gauge-Higgs theory
\bea
     S_L = {\b \over 3} \sum_{p} \text{ReTr}[U(p)] + 
               {\k \over 3} \sum_{x}\sum_{\m=1}^4 \text{Re}\Bigl[\Omega^\dg(x) U_\m(x) \Omega(x+\hat{\m})\Bigr] \ ,
\label{ghiggs}
\eea
where $\Omega(x)$ is a unimodular scalar field $\Omega^\dg(x) \Omega(x)=1$  transforming under gauge transformations
$\Omega(x) \ra g(x) \Omega(x)$ in the fundamental representation.  For $\k\ne 0$ we determine the effective PLA at non-zero
$\m$, and solve the effective theory in the mean-field approximation, postponing more sophisticated methods \cite{Mercado:2012ue,Aarts:2011zn,Fromm:2011qi} to a later study.  We will also determine the PLA corresponding to a lattice SU(3) gauge theory with massive quarks and
large chemical potential, and again apply mean field methods to compute observables.

    In section \ref{method} we will review the relative weights method, and show how the introduction of an imaginary chemical potential, in the gauge-matter system, allows us to determine the $\m$-dependence of the center symmetry-breaking terms.  The PLA corresponding to SU(3) pure-gauge theories at $\b=5.6, 5.7$ on a $16^3 \times 6$ lattice volume is derived in section
\ref{pure}.  The main concern of this article, which is the effective action for a gauge theory coupled to matter fields at finite
chemical potential, is the subject of section \ref{higgs}, where we derive the PLA for the SU(3) gauge-Higgs model \rf{ghiggs} in the confinement-like region, again at $\b=5.6$ on a $16^3 \times 6$ lattice, and for $\k=3.6,3.8,3.9$, which is just below the
crossover to a Higgs-like region. The effective PLA for the gauge-Higgs theory at $\k=3.9$, and the effective PLA for an SU(3) gauge field coupled to massive quarks, are solved in the mean-field approximation, following ref.\ \cite{Greensite:2012xv}, in section \ref{mf}.  We conclude in section \ref{conclude}.
    
\section{\label{method}The relative weights method}

    Let $\cal{U}$ denote the space of all Polyakov line (i.e.\ SU(3) spin) configurations $U_\vx$ on the lattice volume.  Consider any path through this configuration space $U_\vx(\l)$ parametrized by $\l$.  The relative weights method enables us to compute the derivative
of the effective action $S_P$ along the path
\beq
            \left( {d S_P \over d \l}  \right)_{\l=\l_0}
\eeq
at any point $\{U_\vx(\l_0)\} \in \cal{U}$.  By computing appropriate path derivatives, the aim is to determine $S_P$ itself.

    The relative weights method is based on the observation that while the path integral in \rf{S_P}, leading to the Boltzman
weight $e^{S_P}$, may be difficult to compute directly for a particular configuration $U_\vx$, the ratio of such path integrals for slightly different Polyakov line configurations (the ``relative weights'') can be expressed as an expectation value, which can be computed by standard lattice Monte Carlo methods.  Let
\bea
U'_\vx = U_\vx(\l_0 + \oh \D \l)  ~~~,~~~
U''_\vx = U_\vx(\l_0 - \oh \D \l ) ~~~,
\eea
denote two Polyakov line configurations that are nearby in $\cal{U}$,
with $S'_L,S''_L$ the lattice actions with timelike links $U_0(\vx,0)$ on a $t=0$ timeslice held fixed to  $U_0(\vx,0)=U'_\vx$
and $U_0(\vx,0)=U''_\vx$ respectively.  Defining
\bea
           \D S_P = S_P[U'_\vx] - S_P[U''_\vx] \ ,
\eea      
we have from \rf{S_P},
\bea
e^{\D S_P} &=&  {\int  DU_k  D\phi ~  e^{S'_L} \over \int  DU_k  D\phi ~  e^{S''_L} }
\non \\ 
&=& {\int  DU_k  D\phi ~  \exp[S'_L-S''_L] e^{S''_L} \over \int  DU_k  D\phi ~  e^{S''_L} }
\non \\
&=& \Bigl\langle  \exp[S'_L-S''_L] \Bigr\rangle'' \ ,
\eea
where $\langle ... \rangle''$ indicates that the VEV is to be taken in the probability measure
\bea
{e^{S''_L} \over  \int  DU_k  D\phi ~  e^{S''_L} } \ .
\eea
Then 
\beq
           \left( {d S_P \over d \l}\right)_{\l=\l_0} \approx {\D S_P \over \D \l} \ .
\eeq  
We are therefore able to compute numerically the derivative of the true effective action $S_P$ along any path in configuration space.  The problem is to choose path derivatives which will enable us to deduce $S_P$ itself.

\subsection{Symmetries of $\mathbf S_P$}

    The PLA $S_P$ inherits, from the underlying gauge theory, an invariance under local transformations
\beq
             U_\vx \ra  g_\vx U_\vx g^\dg_\vx \ ,
\label{symmetry}
\eeq
where $g_\vx$ is a position-dependent element of the SU($N$) group.  This means that $S_P$ can depend on holonomies
only through local traces of powers of holonomies $\tr[U^p_\vx]$; there can be no dependence on expressions such as $\tr[U_\vx U_\vy]$, since for
$\vx \ne \vy$ this term is not invariant under \rf{symmetry}.  Equivalently, the invariance \rf{symmetry} means that $S_P$ depends only on the eigenvalues of the holonomies $U_\vx$.  We take the term  ``Polyakov line'' in an SU($N$) theory to refer to the trace of the Polyakov line holonomy
\beq
             P_\vx \equiv  {1 \over N} \tr[U_\vx] \ .
\label{P}
\eeq
The SU(2) and SU(3) groups are special in the sense that $P_\vx$ contains enough information to determine the eigenvalues
of $U_\vx$ providing, in the SU(3) case, that $P_\vx$ lies in a certain region of the complex plane.  Explicitly, if we denote the 
eigenvalues of $U_\vx$ as $\{e^{i\th_1},e^{i\th_2},e^{-i(\th_1+\th_2)}\}$, then $\th_1,\th_2$ are determined by separating \rf{P} into its real and imaginary parts, and solving the resulting transcendental equations 
\bea
  \cos(\th_1) + \cos(\th_2) + \cos(\th_1+\th_2) &=& 3 \text{Re}[P_x] \ ,
\non \\
  \sin(\th_1) + \sin(\th_2) - \sin(\th_1+\th_2) &=& 3 \text{Im}[P_x] \ .
\label{eigenvalues}
\eea
In this sense the PLA for SU(2) and SU(3) lattice gauge theories at $\m=0$ is a function of only the Polyakov lines 
$P_\vx$.\footnote{For $N>3$ colors, reconstruction of the eigenvalues would require traces of higher powers of the holonomy.  The present article is concerned specifically with the SU(3) gauge group, and the possible generalization of our procedure to larger gauge groups will not be considered here.}

   In a pure-gauge SU(N) theory, or in an SU(N) gauge theory with matter fields in zero $N$-ality representations of the gauge group, there is a sharp distinction between the confinement and deconfinement phases, based on whether or not the invariance with respect to global center symmetry is spontaneously broken.  In the confinement phase, this means that the SU(3) 
PLA $S_P$ must also be invariant under global transformations $P_\vx \ra z P_\vx$, where $z$ is an element of the center subgroup $Z_3$.  Center symmetric actions are also independent of chemical potential, introduced via $e^\m,e^{-\m}$ factors in the $U_0$ and $U_0^\dg$.  Only terms in the action which explicitly break center symmetry will depend on the chemical potential introduced in this way, and pass on that dependence, along with explicit center symmetry-breaking, to the effective action.   

    Motivated by our previous work on the PLA of SU(2) lattice gauge theory \cite{Greensite:2013yd,Greensite:2013bya}, we will  focus on the Fourier (or ``momentum'') components $a_\vk = a^R_\vk + i a^I_\vk$ of Polyakov line configurations, where
\beq
             P_\vx = \sum_\vk  a_\vk e^{i \vk \cdot \vx} \ ,
\eeq
and compute via relative weights the path derivatives with respect to the real part of $a_\vk$
\beq
           O_\vk(\a) = {1\over L^3}\left( {\pa S_P \over \pa a^R_{\vk}}\right)_{a_\vk = \a} \ ,
\label{O}
\eeq
where $L$ is the extension of the cubic lattice and $\a$ is real.  We will see below that $O_\vk$ has a simple dependence on the lattice momentum $k_L$, where
\beq
           k_L = 2 \sqrt{ \sum_{i=1}^3 \sin^2(k_i/2)} \ ,
\eeq
and can be used to determine $S_P$, at least up to terms bilinear in the Polyakov lines.

\subsection{\label{imag}Use of the imaginary chemical potential}

    In the confinement phase of a pure gauge theory, the part of $S_P$ which is bilinear in $P_\vx$ is constrained by center symmetry to a single term of the form
\beq
             S_P = \sum_{\vx \vy} P_\vx P^\dg_\vy K(\vx-\vy) \ .
\label{pureS}
\eeq
In the presence of matter fields which break the center symmetry, other terms proportional to
\beq
\sum_\vx (P_\vx + P^\dg_\vx) ~~,~~ \sum_\vx (P^2_\vx + P^{2\dg}_\vx) ~~,~~
    \sum_{\vx \vy} (P_\vx P_\vy + P^\dg_\vx P^\dg_\vy) Q(\vx-\vy) \ ,
\eeq
will appear in $S_P$ at the bilinear level.  Now $S_P$ at finite chemical potential $\m$ is given by the change of variables
shown in \rf{convert}, so one might naively imagine that these symmetry breaking terms would convert to
\beq
\sum_\vx (P_\vx e^{\m/T} + P^\dg_\vx e^{-\m/T}) ~~,~~ \sum_\vx (P^2_\vx e^{2\m/T} + P^{2\dg}_\vx e^{-2\m/T}) ~~,~~
    \sum_{\vx \vy} (P_\vx P_\vy e^{2\m/T} + P^\dg_\vx P^\dg_\vy e^{-2\m/T}) Q(\vx-\vy) \ ,
\eeq
i.e.\ that terms linear in $P_\vx,P^\dg_\vx$ are proportional to $e^{\m/T}$ and $e^{-\m/T}$, respectively, while terms quadratic in $P$ or $P^\dg$ are proportional to $e^{2\m/T}$ or $e^{-2\m/T}$.  But this is a little too simple.  Going back to the Polyakov line holonomies, we see that $S_P$ might contain, e.g., center symmetry-breaking
terms such as
\beq
    c_1 \sum_\vx (\tr U_\vx + \tr U^{\dg}_\vx) + c_2 \sum_\vx (\tr U_\vx^2 + \tr U^{\dg 2}_\vx) \ .
\eeq
Under the transformation
\beq
          U_\vx \ra e^{\m/T} U_\vx ~~ , ~~ U^\dg \ra e^{-\m/T} U^\dg
\label{transform}
\eeq
these would go over to
\beq
 c_1 \sum_\vx (\tr U_\vx e^{\m/T} + \tr U^{\dg}_\vx e^{-\m/T}) + c_2 \sum_\vx (\tr U_\vx^2 e^{2\m/T}
 + \tr U^{\dg 2}_\vx e^{-2\m/T}) \ .
\label{example1}
\eeq
Now we apply the SU(3) group identities
\beq
\tr[U_\vx^2] = 9 P_\vx^2 - 6 P^\dg_\vx  ~~~,~~~ \tr[U_\vx^{\dg 2}] = 9 P_\vx^{\dg 2} - 6 P_\vx \ ,
\label{identities}
\eeq
and obtain
\beq
\sum_\vx \Bigl\{ (3c_1 e^{\m/T} - 6c_2 e^{-2\m/T}) P_\vx + (3c_1 e^{-\m/T} - 6c_2 e^{2\m/T}) P^\dg_\vx \Bigl\} 
   + 9c_2 \sum_\vx (P_\vx^2 e^{2\m/T}+ P^{\dg 2}_\vx e^{-2\m/T}) \ .
\eeq
If we would reverse the order of operations, first applying the SU(3) group identities \rf{identities} and then the transformation
\rf{transform}, we would have instead
\beq
\sum_\vx \Bigl\{ (3c_1 - 6c_2) e^{\m/T} P_\vx + (3c_1 - 6c_2) e^{-\m/T} P^\dg_\vx \Bigl\} 
   + 9c_2 \sum_\vx (P_\vx^2 e^{2\m/T}+ P^{\dg 2}_\vx e^{-2\m/T}) \ .
\eeq
It follows that if we only knew the effective action at $\m=0$ in powers of $P_x$,  rather than directly in terms of holonomies,  then the naive application of \rf{transform} would lead to the wrong answer at $\m \ne 0$.

   This problem was raised, and a solution was proposed, already in ref.\ \cite{Greensite:2012xv}.  The idea is to carry out the
relative weights calculation in a lattice gauge theory with an imaginary chemical potential $\mu/T=i\th$.   This is done by simply
multiplying the fixed configurations $U'_\vx,U''_\vx$ of timelike links at $t=0$ by an $\vx$-independent phase factor 
$e^{i\th}$, and calculating the path derivatives of $S_P$ at each $\th$ of a set of $\th$ values. This enables us to
separate, in the path derivatives $O_{\vk}(\a,\th)$, terms which are $\th$-independent from terms which depend on 
$\cos(\th), \cos(2\th)$ and so on.  From knowledge of the $\th$-dependence, we are able to work out the $\m$-dependence of the various terms in $S_P$.  This procedure will be illustrated in detail in section \ref{higgs} below.

\subsection{Background momentum modes}

    Since we are computing derivatives of $S_P$ with respect to individual momentum components, there is a question about the other momentum modes which are not differentiated.  Suppose we are differentiating with respect to the Fourier component
$a_\vk$.  Should the other components $a_{\vq \ne \vk}$ be set to zero, or to something else?
    
    There is clearly a danger in setting all other $a_\vq=0$.  This means that we are computing the path derivative in a highly atypical region of configuration space, a region which contributes essentially nothing to the partition function.  For the purpose of 
determining $S_P$, it is safer to carry out the calculation in a region of $\cal{U}$ which has the optimum ``energy-entropy"
balance, and which provides the typical thermalized configurations found in a Monte Carlo simulation.  Ideally, then, we would 
like to carry out the calculation of the path derivative $O_\vk(\a)$ precisely at a configuration in $\cal{U}$ which is generated by
the lattice Monte Carlo method.

    This ideal is only attainable in the large volume, $\a \ra 0$ limit.  In practice our procedure is as follows:  We first run a standard Monte Carlo simulation, generate a configuration of Polyakov line holonomies  $U_\vx$, and compute the Polyakov lines $P_\vx$.  We then set the momentum mode $a_\vk=0$ in this configuration to zero, to obtain the configuration $\widetilde{P}_\vx$, where
\beq
             \widetilde{P}_\vx =  P_\vx - \left({1\over L^3} \sum_\vy P_\vy e^{-i\vk \cdot \vy}\right) e^{i\vk \cdot \vx} \ .
\eeq
Then define
\bea
            P''_\vx &=& \Bigl(\a - \oh \D \a \Bigr) e^{i\vk \cdot \vx} + f \widetilde{P}_x \ ,
\non \\
            P'_\vx &=& \Bigl(\a + \oh \D \a \Bigr) e^{i\vk \cdot \vx} + f \widetilde{P}_x \ ,
\eea
where $f$ is a constant close to one.  We derive the eigenvalues of the corresponding holonomies $U''_x$ and $U'_x$,
whose traces are $P''_\vx,P'_\vx$ respectively, by solving \rf{eigenvalues}.  The holonomies themselves can be taken to be diagonal matrices, without any loss of generality, thanks to the invariance \rf{symmetry}.  If we could take $f=1$, then in
creating $P''_\vx,P'_\vx$ we are only modifying a single momentum mode of the Polyakov lines of a thermalized
configuration.  However, there are two problems with setting $f=1$.  The first, which already came up in our SU(2)
calculations, is that at $f=1$ and finite $\a$ there are usually some lattice sites where $|P'_\vx|,|P''_\vx| > 1$, which is not allowed.
In SU(3) there is the further problem that at some sites the transcendental equations \rf{eigenvalues} have no solution
for real angles $\th_1,\th_2$.  So we are forced to choose $f$ somewhat less than one; in practice we have used
$f=0.8$.  The choice $f=1$ is only possible in the large volume, $\a \ra 0$ limit.  We have checked that our numerical 
results are insensitive to small changes in $f$.

    From the holonomy configurations $U''_x, U'_x$ we can compute $\pa S_P / \pa a^R_\vk$ by the relative weights method.
This procedure is repeated a number of times (ranging from 30 to 180, depending on the simulation), starting each time from a different thermalized configuration $U_\vx$, and the results for $\pa S_P / \pa a^R_\vk$ are averaged. The standard deviation of the observable $\pa S_P / \pa a^R_\vk$ within a sample of configurations is smallest at the low momenta which dominate the long range behavior of the correlator, and is typically an order of magnitude less than the average value of the correlator at the lowest $k_L$.  As $k_L$ increases and the value of the observable drops towards zero, the standard deviation is eventually on the order of the average value.  Of course, the overall statistical error depends on the sample size.  For the data shown below for $k_L=0$ in Fig.\ \ref{zmode56}, with a sample size of 160, the statistical error is two orders of magnitude smaller than the average values, which is smaller than the symbol size.

\subsection{Limitations of the method}

    Effective actions have, in general, an infinite number of terms, and some truncation is unavoidable.  At finite chemical potential, $S_P$ can be expanded in powers of fugacity
\beq
            S_P = \sum_{s=-\infty}^\infty e^{s \mu/T} S^{(s)}_P[U_\vx,U^\dg_\vx] \ .
\label{expand}            
\eeq
If this is a convergent series (rather than an asymptotic expansion), it implies that $S^{(s)}_P$ must drop off with $s>0$ faster than any exponential of $-s$.  But whether convergent or asymptotic, it is certain that as $\m$ increases one must keep a increasing number of terms in the sum in order to have an accurate
approximation to the effective action.  Since these higher terms will be very small in magnitude at zero or imaginary
chemical potential, it is certain that they will be missed, beyond some order in the fugacity, in a relative weights computation.

   In this article we will be able to determine the contributions to $S_P$ up to second order in fugacity, and to second order
in products of the Polyakov line holonomies.   These restrictions are not absolute, and can probably be overcome to some extent by further development of our method.  But it should be clear from the start that we are always bound to miss terms in the sum that will become important at sufficiently large chemical potential.  Hopefully our methods will determine enough of $S_P$ that the interesting transitions in the $\m-T$ phase diagram for light quarks will be accessible, and that the large
particle densities associated with such transitions are obtained at moderate, rather than enormous fugacities.   But this issue can only be decided by investigation, of the sort we initiate here.

\section{\label{pure}Results for pure gauge theory}                

    We consider the effective action $S_P$ corresponding to an underlying pure SU(3) lattice gauge on a $16^3 \times 6$
lattice, at lattice couplings $\b=5.6,5.7$.  For these couplings the gauge theory is in the confinement phase; the deconfinement transition at $N_t=6$ lattice spacings in the time direction is at $\b=5.89$.

    For a pure SU(3) gauge theory the bilinear form of the effective action is particularly simple, as already noted above.  Expressing \rf{pureS} in momentum components, we have
\beq
            S_P = \sum_\vk a_k a^*_k \tK(\vk) \ ,
\eeq
where 
\beq
             K(\vx-\vy) = {1\over L^3} \sum_\vk \tK(k) e^{-\vk \cdot (\vx-\vy)} \ .
\eeq
We see that for real $\a$
\beq
            {1\over L^3}\left( {\pa S_P \over \pa a^R_{\vk}}\right)_{a_\vk = \a} = 2 \tK(\vk) \a \ .
\label{observable}
\eeq

\begin{figure}[t!]
\centerline{\scalebox{0.6}{\includegraphics{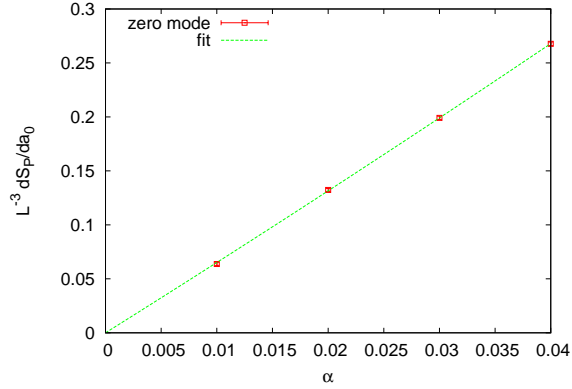}}}
\caption{The path derivative of $S_P$ with respect to the real part of the mode at $k_L=0$,  evaluated at several values 
$a_0=\a$ of the $k_L=0$ mode.  This is for an underlying pure gauge theory at $\b=5.6$. The data is fit to $A\a + B\a^2$, with $A=2\tK(0)$. In this figure, and in all other figures below, the lattice volume of the underlying lattice gauge theory is 
$16^3 \times 6$.}
\label{zmode56}
\end{figure}

    We compute the left hand side at several values of $\a$, and divide each result by $\a$.  The values almost coincide within
errors, apart from the values at $k_L=0$, where there is a small but noticeable ($\sim 3\%$) deviation. For the data point at $k_L=0$ we therefore extrapolate to $\a=0$ by fitting the data to the curve $A\a + B\a^2$, as shown in Fig.\ 
\ref{zmode56}.\footnote{This implies, of course, that there must be terms in $S_P$ which are higher order than quadratic.
We will return to this issue later; for the moment we are concerned with computing only the bilinear terms.}
Then $2\tK(0)=A$ is the extrapolated value.   

\begin{figure}[t]
\subfigure[]  
{   
 \label{G56}
 \includegraphics[scale=0.6]{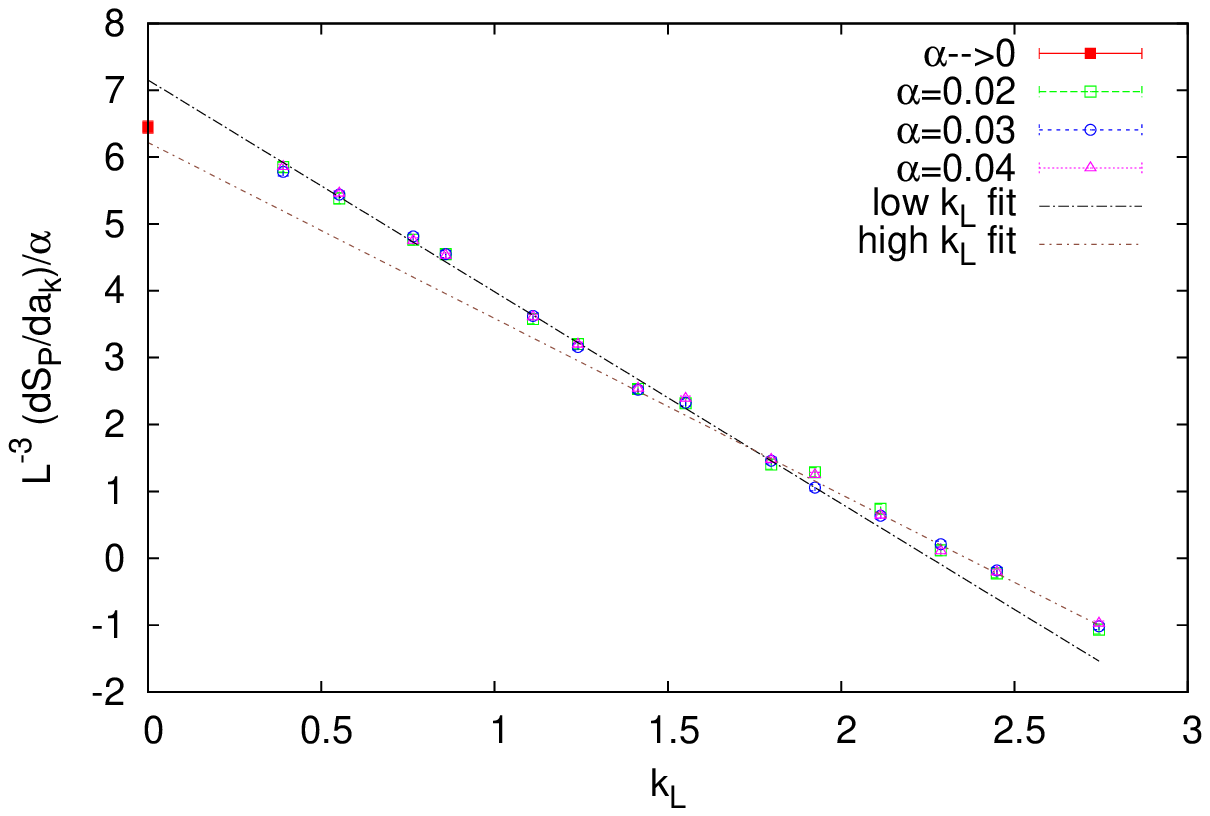}
}
\subfigure[]  
{   
 \label{kmat}
 \includegraphics[scale=0.6]{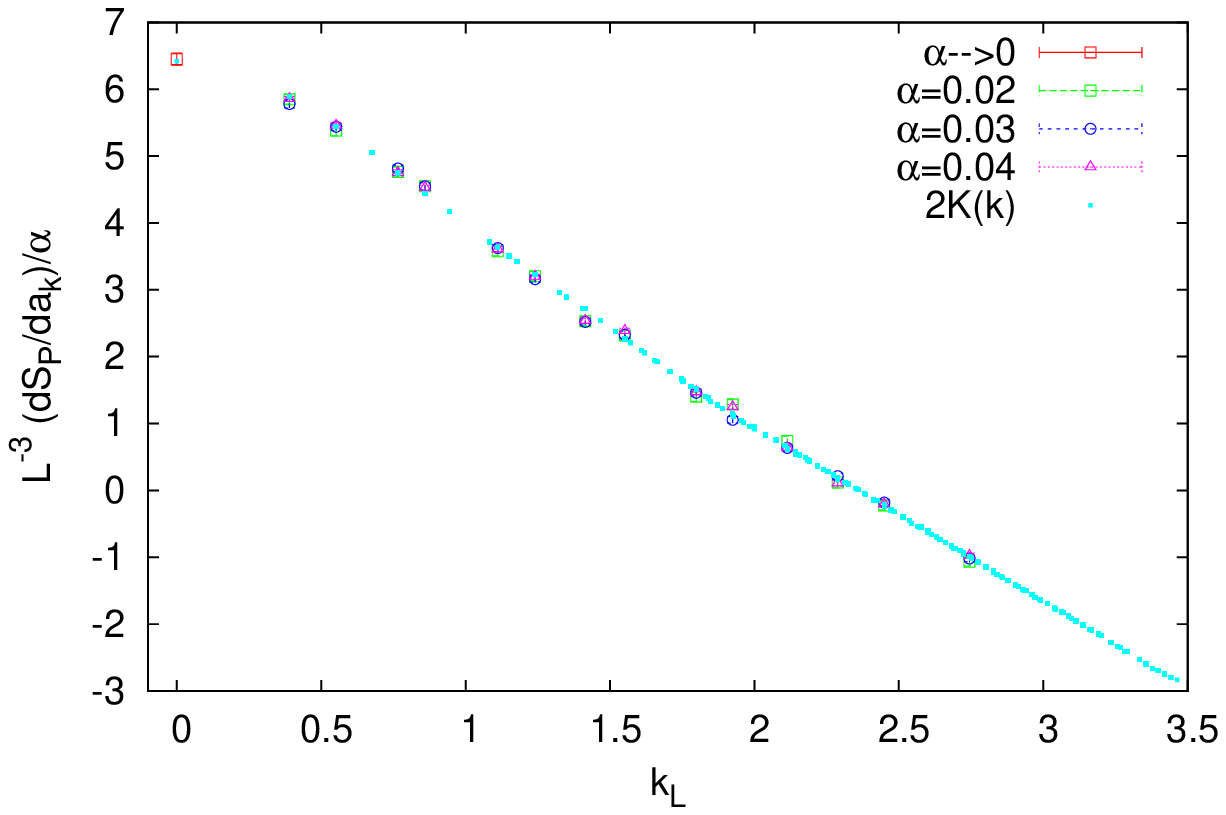}
}
\caption{Path derivatives of $S_P$ with respect to momentum modes $a_\vk$, evaluated at $a_\vk=\a$ and then
divided by $\a$, for 15 values of $k_L$. The rescaled derivatives are shown for several values of $\a$, with the exception of
the point at $k_L=0$, which is the value determined from the data in Fig.\ \ref{zmode56}.  This is for an underlying pure gauge theory at $\b=5.6$. (a) data points fit by two straight lines. (b) the data points together
with $2\tK(k_L)$, determined by the procedure explained in the text. }
\label{b56data}
\end{figure}

\begin{figure}[htb]
\centerline{\scalebox{0.9}{\includegraphics{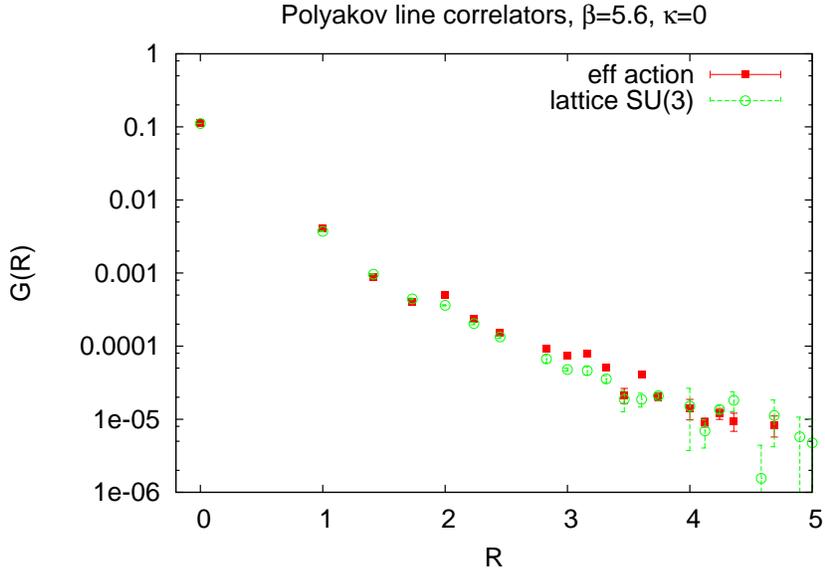}}}
\caption{The Polyakov line correlators for pure gauge theory at $\b=5.6$,  computed from numerical simulation of the 
effective PLA $S_P$, and from simulation of the underlying lattice SU(3) gauge theory.}
\label{corr56}
\end{figure} 

    The data for   
\beq
{1\over \a} {1\over L^3}\left( {\pa S_P \over \pa a^R_{\vk}}\right)_{a_\vk = \a}
\label{dS_over_alpha}
\eeq
at all $k_L$ is displayed in Fig.\ \ref{G56}, together with the value at $k_L=0$ extrapolated to $\a=0$. In this and all other graphs with $k_L$ on the $x$-axis we have used momenta $\vk$ with components $k_i=2\pi m_i/L$ ($L=16$ in this case), for the following triplets $\vec{m}=(m_1 m_2 m_3)$ of mode numbers:
\bea
& & (000),(100), (110),(200),(210),(300),(311),(400),(322),(430),(333),
\non \\
& & (433),(443),(444),(554) \ .
\eea
The main point to notice in Fig.\  \ref{G56} is that most of the data fits on a straight line, with the exception of the point at $k_L=0$. This was also what we found for SU(2) gauge theory in our previous work \cite{Greensite:2013yd,Greensite:2013bya}: the very low momentum data tends to bend away from a straight-line fit.   There are no indications of rotational symmetry breaking that might arise due to the cubic lattice.
A new feature that has turned up in the SU(3) case is that the higher momentum points, at $k_L \ge k_0 \approx 1.8$, seem to fit a straight line with a slightly different slope than the line which fits the $k_L <k_0$ data.  This change of slope will be more pronounced in the further examples below.  

   So the data seems to depend only on $k_L$, and fits a straight line in the ranges $k_{min} < k_L < k_0$, and $k_L > k_0$,  where $k_0 \approx 1.8$ is the point where the slope suddenly changes, and $k_{min}=0$ .  We therefore
write the kernel as a function of just $k_L$, rather than the wavevector $\vk$.  The way that we fit the data is to first do a linear fit to $c_1 - 4c_2k_L$ for the data in the range $k_{min} < k_L < k_0$, and a fit to $b_1- 4b_2 k_L$ in the high momentum range 
$k_L > k_0$.  Then set
\bea
          \tK^{fit}(k_L) = \left\{ \begin{array}{cl}
                 \oh c_1 - 2 c_2 k_L & k_L \le k_0 \cr \\
                 \oh b_1 - 2 b_2 k_L & k_L > k_0 \end{array} \right. \ .
\eea
Next define the position-space kernel with a long distance cutoff $r_{max}$
\beq
    K(\vx-\vy) = \left\{ \begin{array}{cl}
                   {1\over L^3}\sum_\vk \tK^{fit}(k_L) e^{i\vk\cdot (\vx-\vy)} & |\vx-\vy| \le r_{max} \cr \\
                      0 & |\vx-\vy| > r_{max} \end{array} \right. \ .
\eeq
The cutoff $r_{max}$ is chosen so that, upon transforming {\it this} kernel back to momentum space, the resulting $\tK(k)$ also fits the low-momentum data at $k_L \le k_{min}$, where $k_{min}=0$ in this example.  The procedure is described in more detail in \cite{Greensite:2013bya}. The point $k_0$ is determining by carrying out two straight line fits to the data in the regions 
$k_{min}<k_L\le k_0$ and $k_L \ge k_0$, and then checking that the two straight lines intersect at $k_0$.  We vary $k_0$ until this matching condition is satisfied. The quantity $2\tK(k_L)$ obtained by this method is shown in Fig.\ \ref{kmat}, together with the data for \rf{dS_over_alpha}. 

   Once again, this is all very similar to our previous findings for the SU(2) PLA.  The only difference is that we now have to
allow for a different linear fit for higher momentum points, in this case for $k_L > 1.8$.  The physical mechanism behind this
abrupt change in slope at $k_L=k_0$ is not yet clear to us.

    Now that we have obtained the kernel $K(\vx-\vy)$ we can simulate the effective PLA, which is an SU(3) spin model 
\rf{pureS}, by standard lattice Monte Carlo methods, and calculate the spin-spin correlator \rf{GR}.  We can compare this
with the corresponding Polyakov line correlator computed in the underlying SU(3) lattice pure gauge theory, at $\b=5.6$
on a $16^3 \times 6$ lattice volume.  The comparison (including off-axis separations) is shown in Fig.\ \ref{corr56}.
Allowing for the fact that the data is a little noisy beyond $R=4$, this seems like good agreement.

\begin{figure}[htb]
\centerline{\scalebox{0.7}{\includegraphics{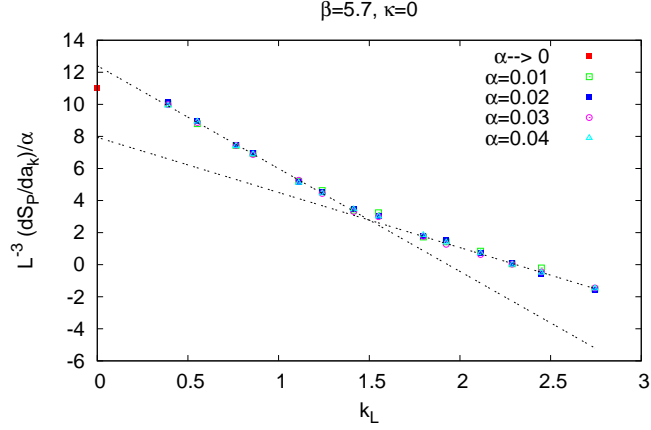}}}
\caption{Same as Fig.\ \ref{G56}, but for the pure gauge theory at $\b=5.7$.}
\label{G57}
\end{figure}

\begin{figure}[htb]
\centerline{\scalebox{0.9}{\includegraphics{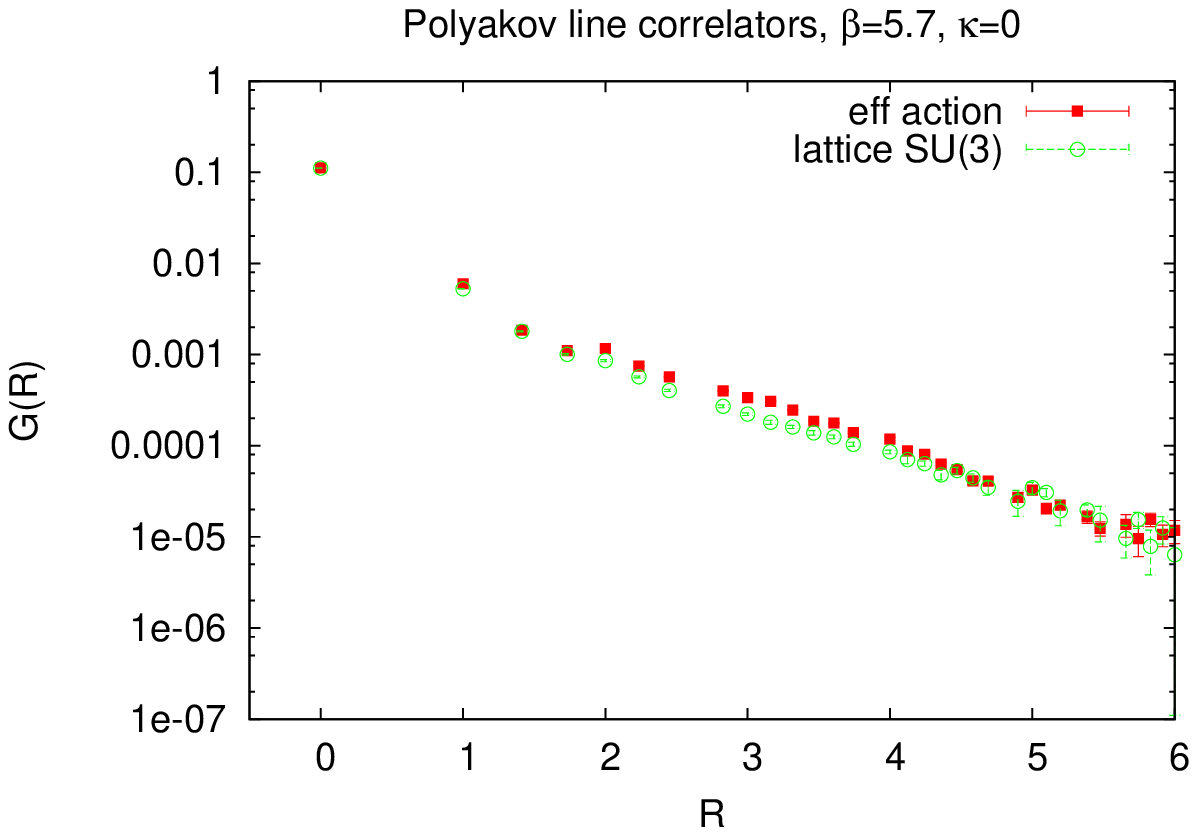}}}
\caption{The Polyakov line correlators for pure gauge theory at $\b=5.7$,  computed from numerical simulation of the 
effective PLA $S_P$, and from simulation of the underlying lattice SU(3) gauge theory.}
\label{corr57}
\end{figure}

    The next example, coming a little closer to the deconfinement transition at $\b=5.89$, is the pure gauge theory at
$\b=5.7$.  We again calculate the observable \rf{dS_over_alpha} at several $\a$ values, and we find again that the data points
overlap, excluding the point at $k_L=0$.  Extrapolating this point to $\a=0$ by the same method as before, we find results 
for \rf{dS_over_alpha} displayed in Fig.\ \ref{G57}.  This time the change in slope is found at $k_0=1.51$.  We determine the kernel $K(\vx-\vy)$ by the procedure outlined above, and simulate the resulting $S_P$.  The comparison of Polyakov line correlators at off-axis separations is shown in Fig.\ \ref{corr57}.

   The parameters which define the effective action \rf{pureS} in these two examples are given in Table \ref{tab1}.  Note the very substantial increase in parameters $c_1,c_2$ as we approach the deconfinement transition.

\begin{table}[t!]
\begin{center}
\begin{tabular}{|c|c|c|c|c|c|c|} \hline
         $ \b $ &  $c_1$ & $c_2$ & $k_0$ &  $b_1$ & $b_2$ & $r_{max}$ \\
\hline
        5.6 &  7.15(5) & 0.79(1)  & 1.79 &6.22(14) & 0.66(1)    & $\sqrt{29}$   \\ 
        5.7 &  12.41(5) & 1.60(1) & 1.51  & 7.94(14) & 0.86(2) & 6 \\          
\hline
\end{tabular}
\caption{Parameters defining the effective Polyakov line action $S_P$ for pure SU(3) lattice gauge theory on
a $16^3 \times 4$ lattice.} 
\label{tab1}
\end{center}
\end{table}

   It should be emphasized that the bilinear action does not imply that the effective action is a free field theory (any more than a non-linear sigma model is a free field theory), and of course there are an infinite number of non-trivial connected $n$-point functions in the theory.  It is not hard to see, in the context of a strong-coupling expansion, how the bilinear action can generate, e.g., a 3-point correlator $\langle P_x P_y P_z \rangle$ in SU(3).  We have computed the two-point Polyakov line correlator simply because it is the simplest thing to measure; $n=3$ point (and higher) correlators are left for future work.

\section{\label{higgs}Results for SU(3) gauge-Higgs theory}

    We now add a scalar matter term, and consider the SU(3) gauge-Higgs theory \rf{ghiggs} at several different values
of $\k$.  There is an extensive literature on the SU(2) version of this theory (see, e.g., Bonati et al.\ \cite{Bonati:2009pf} and
references therein), and it is well known from the work of Fradkin and Shenker \cite{Fradkin:1978dv} and
Osterwalder and Seiler \cite{Osterwalder:1977pc} that there is no complete separation of the phase diagram
into a confining and a deconfining (or ``Higgs'') phase.  This ties in with the fact that there is no local or semi-local gauge-invariant order parameter which would distinguish the two phases.  In some regions of the $\b-\k$ phase diagram, however, there can be either a first-order transition, or a rapid crossover, from a ``confinement-like'' region to a ``Higgs-like'' region.  The confinement-like region is characterized, as in real QCD, by an area-law falloff of Wilson loops (or an exponential drop in the Polyakov-line correlator) up to some string-breaking scale.  In the Higgs-like region the behavior is more like the electroweak
theory, with no string formation (or linear static potential) at any scale.  In the present exploratory study, we are interested mainly in the confinement-like region, and we will work exclusively at the gauge coupling $\b=5.6$ on a $16^3 \times 6$ lattice volume as before.

\begin{figure}[htb]
\centerline{\scalebox{0.4}{\includegraphics{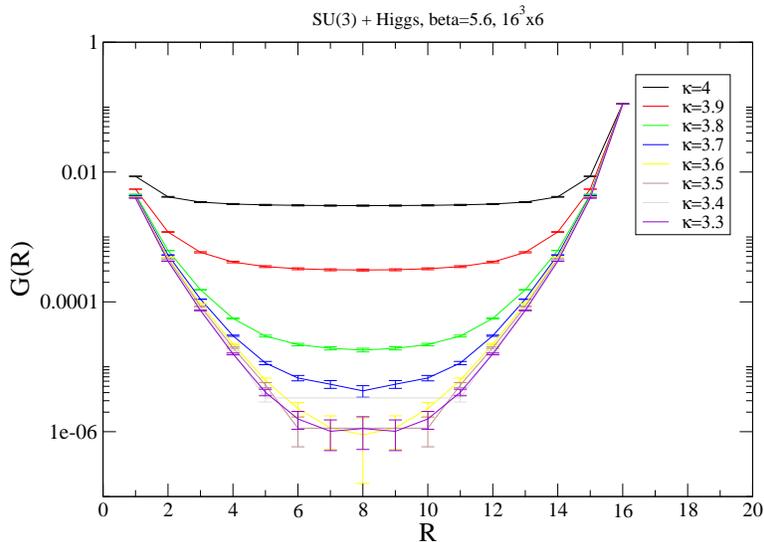}}}
\caption{On-axis Polyakov line correlators computed for the underlying gauge-Higgs theory at $\b=5.6$ and a 
variety of $\k$ values on a $16^3\times 6$ lattice volume.  The correlators have been computed using the L\"uscher-Weisz
noise reduction method.}
\label{pot56kappa}
\end{figure}

   Results for the Polyakov line correlators in the lattice gauge-Higgs theory at a variety of $\k$ values are shown in Fig.\ 
\ref{pot56kappa}.   A calculation of the Polyakov line susceptability does not reveal a phase transition, but there is a peak
in the susceptability at $\k \approx 4$, indicative of a rapid crossover.  Since we are interested in the effects of (relatively) light scalars in the confinement-like regime, we consider $\k$-values close to but just below the crossover, specifically at $\k=3.6,3.8, 3.9$.

    The new feature at $\k > 0$ is that we have to determine the terms in the effective action $S_P$ which explicitly break center symmetry, and also to sort out their behavior at finite chemical potential.  Effective actions which result from integrating out degrees of freedom in the underlying theory will typically involve an infinite number of terms.  Truncation to a finite number of
terms is therefore essential.  We first consider a PLA truncated to terms bilinear in $\tr U_\vx$ and $\tr U^2_\vx$ (and their complex conjugates),  and apply the transform \rf{convert} to obtain the action at finite $\m$.  We then use the identities 
\rf{identities} to express $S_P$ in terms of the Polyakov lines, and finally discard terms involving products of three or more of the $P_\vx$. Even with such a truncation, we will see that some of the terms are negligible, at least until $e^{\m/T}$ is quite large.  So initially we have
\bea
\lefteqn{S_P =} 
\non \\
   &  & \qquad \sum_{\vx \vy} \tr U_\vx \tr U^\dg_\vy K_1(\vx-\vy) + \sum_{\vx \vy} \tr U^2_\vx \tr U^{\dg 2}_\vy K_2(\vx-\vy)
 + a_1 \sum_\vx (\tr U_\vx + \tr U^{\dg}_\vx) + a_2 \sum_\vx (\tr U_\vx^2 + \tr U^{\dg 2}_\vx)
\non \\
&    & \qquad + \sum_{xy} (\tr U_\vx \tr U_\vy + \tr U^{\dg 2}_\vx \tr U^{\dg 2}_\vy) Q_1(\vx-\vy)
  + \sum_{xy} (\tr U^2_\vx \tr U^\dg_\vy + \tr U^{\dg 2}_\vx \tr U_\vy) Q_2(\vx-\vy)
\non \\
&   & \qquad + \sum_{xy} (\tr U^2_\vx \tr U^2_\vy + \tr U^{\dg 2}_\vx \tr U^{\dg 2}_\vy) Q_3(\vx-\vy) \ .
\eea
Then at finite chemical potential, from \rf{convert},
\bea
     \lefteqn{S_P = }
\non \\
& & \qquad \sum_{\vx \vy} \tr U_\vx \tr U^\dg_\vy K_1(\vx-\vy) + \sum_{\vx \vy} \tr U^2_\vx \tr U^{\dg 2}_\vy K_2(\vx-\vy)
 + a_1 \sum_\vx (\tr U_\vx e^{\m/T} + \tr U^{\dg}_\vx e^{-\m/T}) 
\non \\
& &\qquad + a_2 \sum_\vx (\tr U_\vx^2 e^{2\m/T} + \tr U^{\dg 2}_\vx e^{-2\m/T} ) + \sum_{xy} (\tr U_\vx \tr U_\vy e^{2\m/T} + \tr U^{\dg 2}_\vx \tr U^{\dg 2}_\vy e^{-2\m/T}) Q_1(\vx-\vy)
\non \\
& & \qquad + \sum_{xy} (\tr U^2_\vx \tr U^\dg_\vy e^{\m/T} + \tr U^{\dg 2}_\vx \tr U_\vy e^{-\m/T}) Q_2(\vx-\vy)
\non \\
& & \qquad + \sum_{xy} (\tr U^2_\vx \tr U^2_\vy e^{4\m/T} + \tr U^{\dg 2}_\vx \tr U^{\dg 2}_\vy e^{-4\m/T}) Q_3(\vx-\vy)  \ .
\eea
Now apply the identities \rf{identities} to express everything in terms of the Polyakov lines, and discard terms involving a product of three or more lines:
\bea
S_P &=& \sum_{xy} P_\vx P_\vy^\dg K(\vx-\vy) + \sum \Bigl\{(d_1 e^{\m/T} - d_2 e^{-2\m/T}) P_\vx + (d_1 e^{-\m/T} - d_2 e^{2\m/T}) P^\dg_\vx \Bigr\}
\non \\
& & + \sum_{xy} (P_\vx P_\vy Q(\vx-\vy,\mu) +  P^\dg_\vx P^\dg_\vy Q(\vx-\vy;-\m) ) \ ,
\label{SP_for_ghiggs}
\eea
where $d_1 = 9 a_1, ~ d_2 = 6 a_2$; and
\bea
K(\vx-\vy) &=& 9 K_1(\vx-\vy) + 36 K_2(\vx-\vy) \ ,
\non \\
Q(\vx-\vy; \m)  &=& Q^{(1)}(\vx-\vy) e^{-\m/T} + Q^{(2)}(\vx-\vy) e^{2\m/T} + Q^{(4)}(\vx-\vy) e^{-4\m/T} \ ,
\label{Qmu}
\eea
where
\bea
 Q^{(1)}(\vx-\vy) &=& - 18 Q_2(\vx-\vy) ~~,~~ Q^{(2)}(\vx-\vy) = 9 a_2 \d_{\vx \vy} + 9 Q_1(\vx-\vy) ~~,
\non \\   
      Q^{(4)}(\vx-\vy) &=& 36 Q_3(\vx-\vy)  \ .
\eea

    The problem is to determine the kernels $K(\vx-\vy), Q(\vx-\vy;\m)$ and the constants $d_1,d_2$.  For this purpose it is
useful to introduce an imaginary chemical potential $\m/T=i\th$, as discussed in section \ref{imag}.  In momentum space the bilinear action becomes
\bea 
{1\over L^3} S_P &=& \sum_\vk a_\vk a^*_\vk \tK(k_L)
       + a_0\Bigl(d_1 e^{i\th} - d_2 e^{-2i\th} \Bigr) +  a^*_0\Bigl(d_1 e^{-i\th} - d_2 e^{2i\th} \Bigr)
\non \\
& & + \sum_\vk \Bigl(a_\vk a_{-\vk}\tQ(k_L,\m)+ a^*_\vk a^*_{-\vk} \tQ(k_L,-\m)\Bigl) \ .
\eea
Taking the derivative with respect to $a^R_0$, evaluated at $a_0=a^*_0=\a$, we have
\beq
{1\over L^3} \left({\pa S_P \over \pa a^R_0}\right)_{a_0=\a} = 2\tK(0) \a + (2d_1 + 4\tQ^{(1)}(0)\a) \cos(\th)
   - (2d_2 - 4\tQ^{(2)}(0)\a)\cos(2\th) \ .
\eeq
Fitting the data to
\beq
{1\over L^3} \left({\pa S_P \over \pa a^R_0}\right)_{a^R_0=\a} = A(\a) + B(\a) \cos(\th) - C(\a) \cos(2\th)
\label{dSzero_mode}
\eeq
allows us to determine
\beq
\tK(0) = \oh {dA \over d\a} ~~,~~ d_1=\oh B(0) ~~,~~ \tQ^{(1)}(0) = \oq {dB \over d\a} ~~,~~ 
          d_2=\oh C(0) ~~,~~ \tQ^{(2)}(0) = -\oq {dC \over d\a} \ .
\label{from_0_modes}
\eeq
For $\vk \ne 0$, the derivative wrt $a_\vk$ has terms proportional to $a_{-\vk}$.  We set $a_{-\vk}$ to some
constant real value $a_{-\vk}=\s$.  Then
\beq
{1\over L^3} \left({\pa S_P \over \pa a^R_\vk}\right)^{\a_{-\vk}=\s}_{a_\vk=\a} = 2 \tK(k_L) \a + 
  4 (\tQ^{(1)}(k_L) \cos(\th) + \tQ^{(2)}(k_L) \cos(2\th) + \tQ^{(4)}(k_L) \cos(4\th)) \s \ .
\label{Qk}
\eeq
First, setting $\s=0$, we have
\beq
\tK(k_L) = {1\over 2 L^3} {d\over d\a} \left({\pa S_P \over \pa a^R_\vk}\right)^{a_{-\vk}=0}_{a_\vk=\a} \ .
\label{from_k_modes}
\eeq
Then, at small but finite $\s$, we can determine the $\tQ^{(n)}(k_L)$ from the $\th$-dependence of the data.  

\subsection{$\mathbf \k=3.9$}

\begin{figure}[htb]
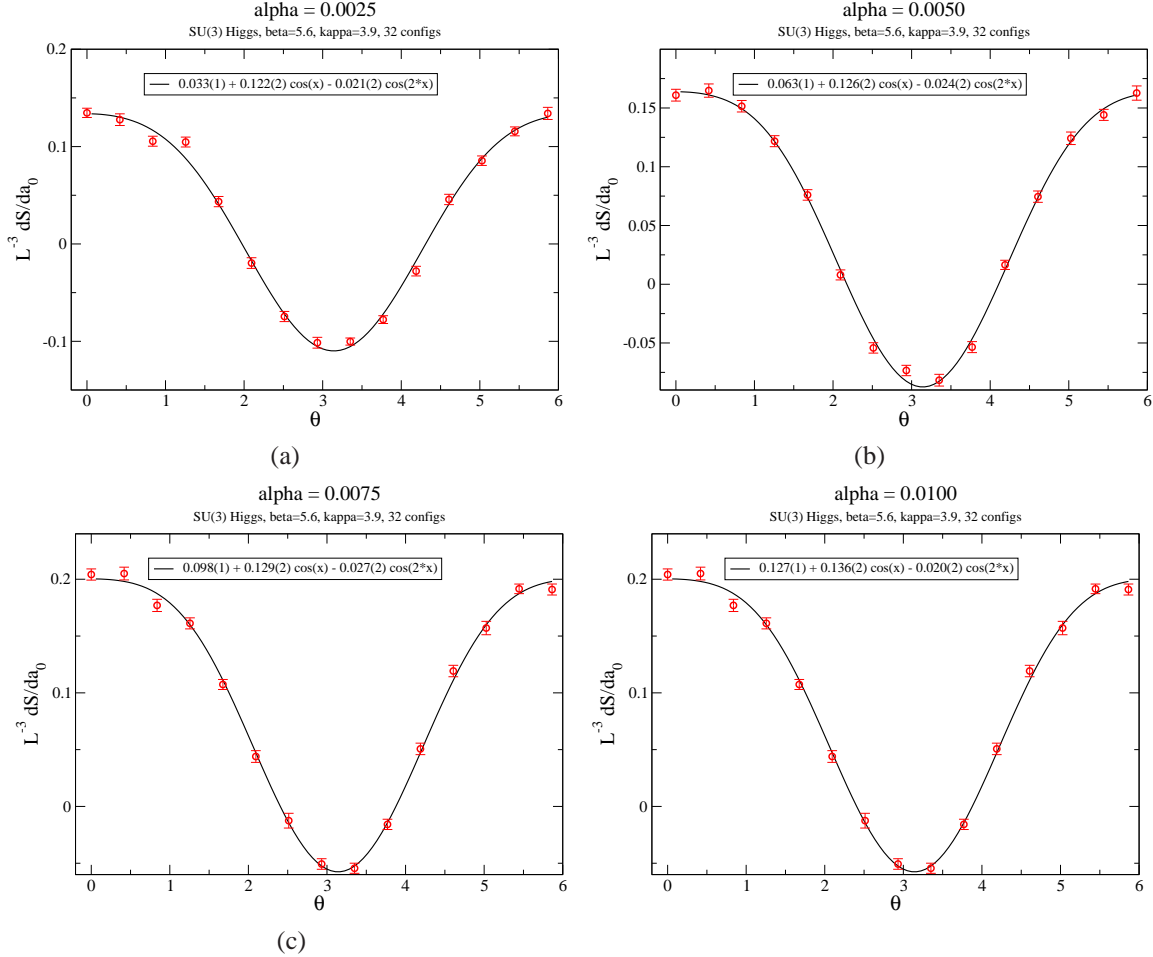

\subfigure[]  
{   
 \label{}
 \includegraphics[scale=0.3]{mu_39_025.eps}
}
\subfigure[]  
{   
 \label{}
 \includegraphics[scale=0.3]{mu_39_050.eps}
}
\subfigure[]  
{   
 \label{}
 \includegraphics[scale=0.3]{mu_39_075.eps}
}
{   
 \label{}
 \includegraphics[scale=0.3]{mu_39_100.eps}
}
\caption{A plot of $L^{-3} \pa S_P/\pa a_0^R$ evaluated at $a_0=\a$, plotted against the imaginary chemical
potential $\mu/T=i\th$.  The data is fit to a truncated cosine series \rf{dSzero_mode} to determine center symmetry-breaking
terms. (a) $\a$=0.0025, (b) $\a$=0.005, (c) $\a$=0.0075, (d) $\a$=0.01.}
\label{muk39}
\end{figure}

\begin{figure}[htb]
\subfigure[]  
{   
 \label{A}
 \includegraphics[scale=0.6]{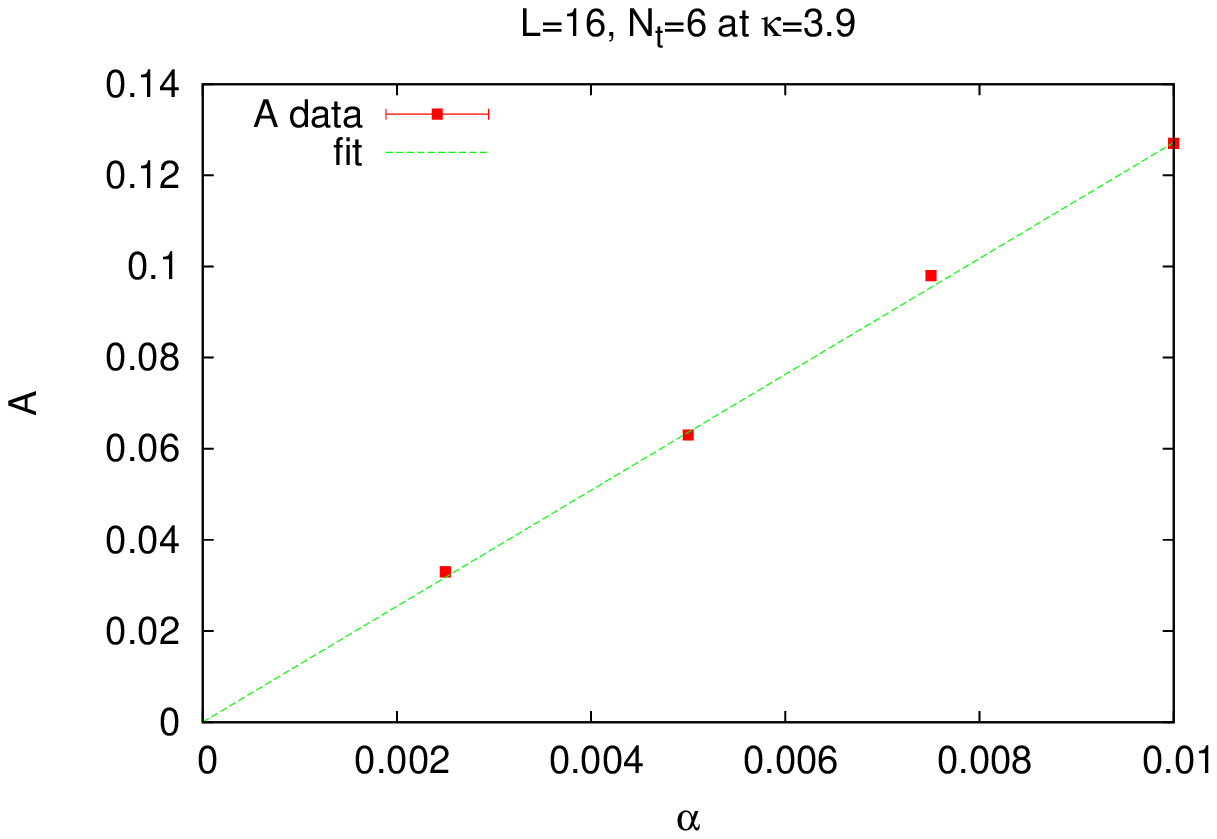}
}
\subfigure[]  
{   
 \label{B}
 \includegraphics[scale=0.6]{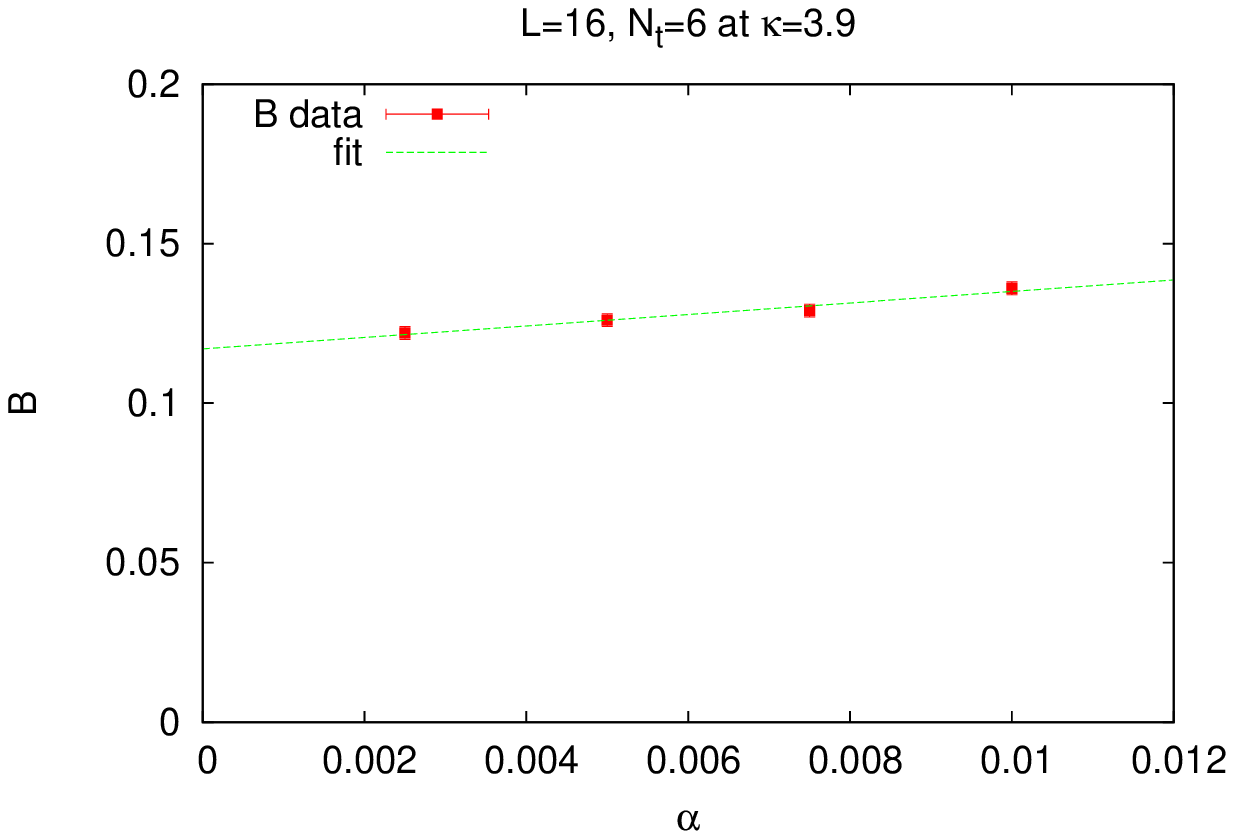}
}
\subfigure[]  
{   
 \label{C}
 \includegraphics[scale=0.6]{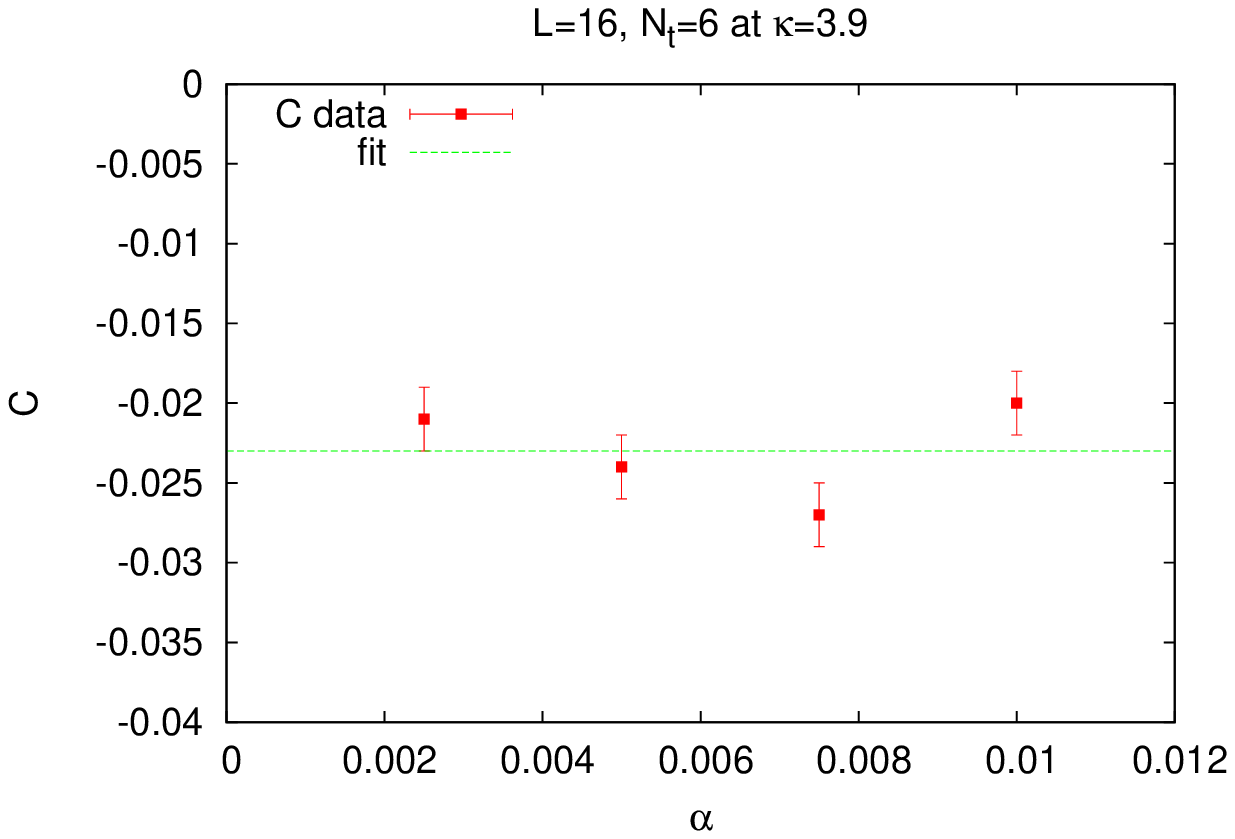}
}
\caption{Coefficients $A,B,C$ of the best fit to the data in Fig.\ \ref{muk39} by a truncated cosine series 
$A+B \cos(\th)+C\cos(2\th)$; these coefficients are displayed
in subfigures (a), (b), and (c) respectively.  The coefficients are
computed at several values of $\a_0=\a$, and the lines shown are a best linear fit.  From the slope and $y$-axis intercept of these lines, we are able to compute parameters of the center-symmetry breaking terms, as explained in the text.}
\label{ABC}
\end{figure}

   We begin by computing the derivative of $S_P$ with respect to the zero-mode $a^R_0$ at 15 values of the imaginary
chemical potential in the range $0\le \th < 2\pi$, and four values of $\a$.  At each $\a$ we fit the results to a truncated cosine series \rf{dSzero_mode}.  The data and the fits are shown in Fig.\ \ref{muk39}.   We then plot $A(\a),B(\a),C(\a)$ extracted
from the cosine fits, and make a linear best fit to the results for $A,B,C$ vs.\ $\a$, as displayed in Fig.\ \ref{ABC}.  From the slope of the best fit lines we get the $\a$-derivatives of these quantities, and the $y$-intercept gives us the values of $A,B,C$
extrapolated to $\a=0$.  The $\a$-derivatives and $\a=0$ values give us $\tK(0),\tQ^{(1,2)}(0),d_1,d_2$, as explained above.

\begin{figure}[htb]
\centerline{\scalebox{0.8}{\includegraphics{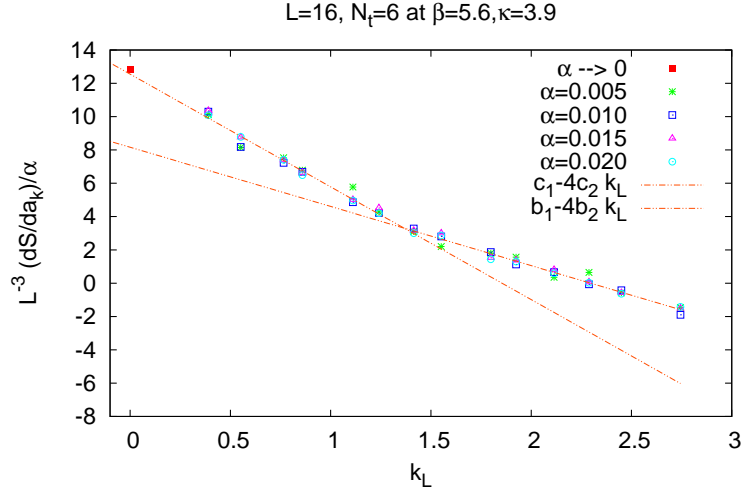}}}
\caption{Path derivatives of $S_P$ with respect to momentum modes $a^R_\vk$, evaluated at $a_\vk=a^*_\vk=\a$ and then divided by $\a L^3$, for 15 values of $k_L$.  This is for an underlying lattice gauge-Higgs theory with $\b=5.6,\k=3.9$.  Data
points at $k_L$ below and above $k_L=1.36$ fall on two straight lines, with a different slope for each line.}
\label{ds}
\end{figure}

    Next we compute the $a_\vk$ derivatives at $\vk \ne 0$ with $\s=a_{-\vk}$ set to zero.  This result, together with our
usual fit by two straight lines, is shown in Fig.\ \ref{ds}.  In this case it appears that the extrapolated $\a\ra 0$ value of $2\tK(0)$
falls very near the $y$-intercept of the first straight line.  That means that we do not see a long-distance cutoff for the
position-space kernel $K(\vx-\vy)$, at least on a $16^3 \times 6$ lattice, and on a lattice volume of this size every point
is coupled to every other point in $S_P$.  This all-points-to-all-points coupling makes the numerical simulation of the effective action a little more time-consuming than before (unless we just truncate the long-distance coupling by hand), but it 
is still possible.

\begin{figure}[htb]
\subfigure[]  
{   
 \label{Q100}
 \includegraphics[scale=0.27]{sigma100.eps}
}
\subfigure[]  
{   
 \label{GQ}
 \includegraphics[scale=0.6]{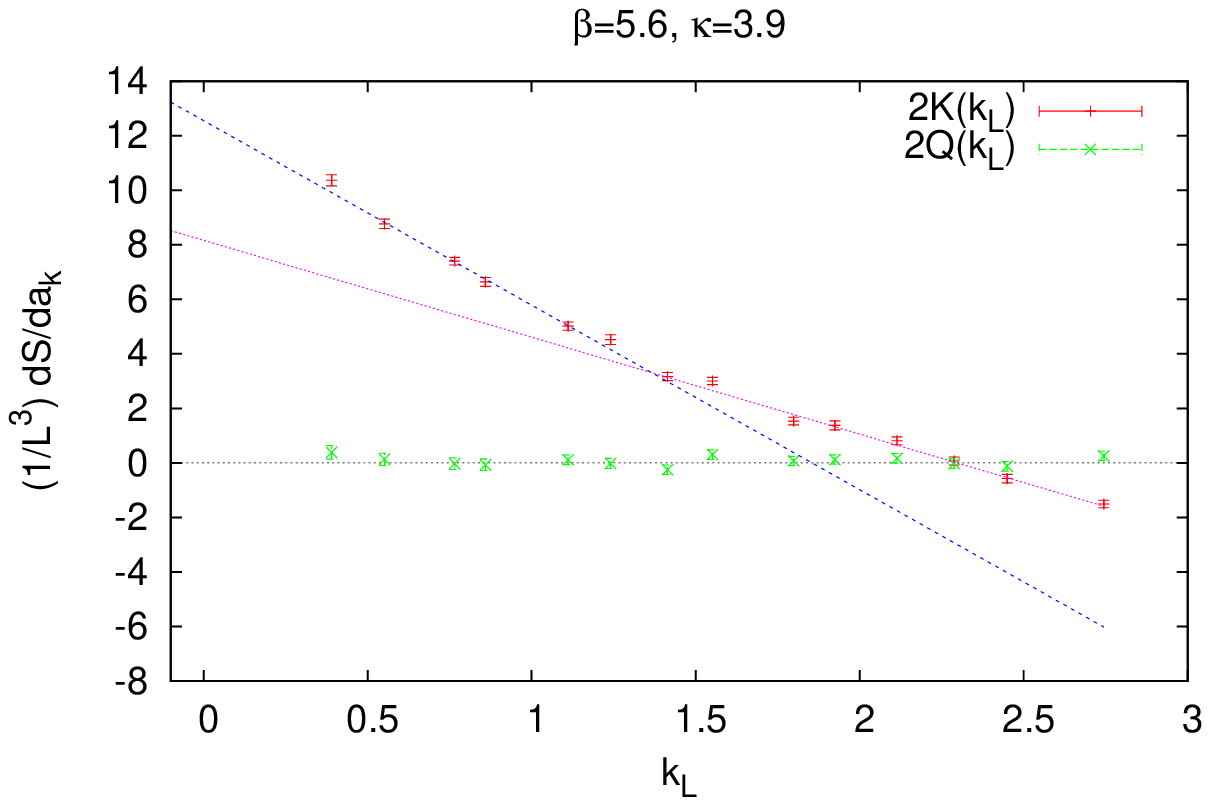}
}
\caption{(a) A plot of the path derivative data vs.\ imaginary chemical potential, analogous to Fig.\ \ref{muk39}, but this time with the derivative taken with respect to the (100) momentum mode at $\a=\s=0.01$.  Statistics are not good enough to determine the coefficient of the $\cos(2\th)$ term.  From data of this sort, taken over a range of $k_L$, we can
in principle determine the semi-local kernel $Q(\vx-\vy,\m)$ of the center-symmetry breaking term involving a product
of Polyakov line variables.  (b) a rough estimate of $2\tQ^{(1)}(k_L)$ vs. $k_L$, shown in comparison with $2\tK(k_L)$.}
\label{Q}
\end{figure}

   Finally we consider $\tQ(k_L,\m)$, with $\m/T=i\th$.   Let us concentrate on the lowest non-zero momentum with components $k_i = 2\pi m_i/L$, with the mode number triplet $(m_1 m_2 m_3)=(100)$, and compute \rf{Qk} at 15 values of $\th$, with $\a=\s=0.01$.  The error bars are large but what we find, seen in Fig.\ \ref{Q100}, is that the $\th$-dependence seems to be dominated by a term proportional to $\cos(\th)$.   However, $\tQ(k_L,\m)$ itself is almost negligible compared to $\tK(k_L)$, as seen in Fig.\ \ref{GQ}, where we plot a rough estimate of $2\tQ^{(1)}(k_L)$ vs.\ $k_L$, based on only three $\th$ values at each
$k_L$. Certainly $\tQ(k_L,\m)$ will become important at sufficiently large and real $\m$ such that $e^\m > 10$, but its contribution at $\m=0$ can be ignored.  

\begin{figure}[htb]
\centerline{\scalebox{0.9}{\includegraphics{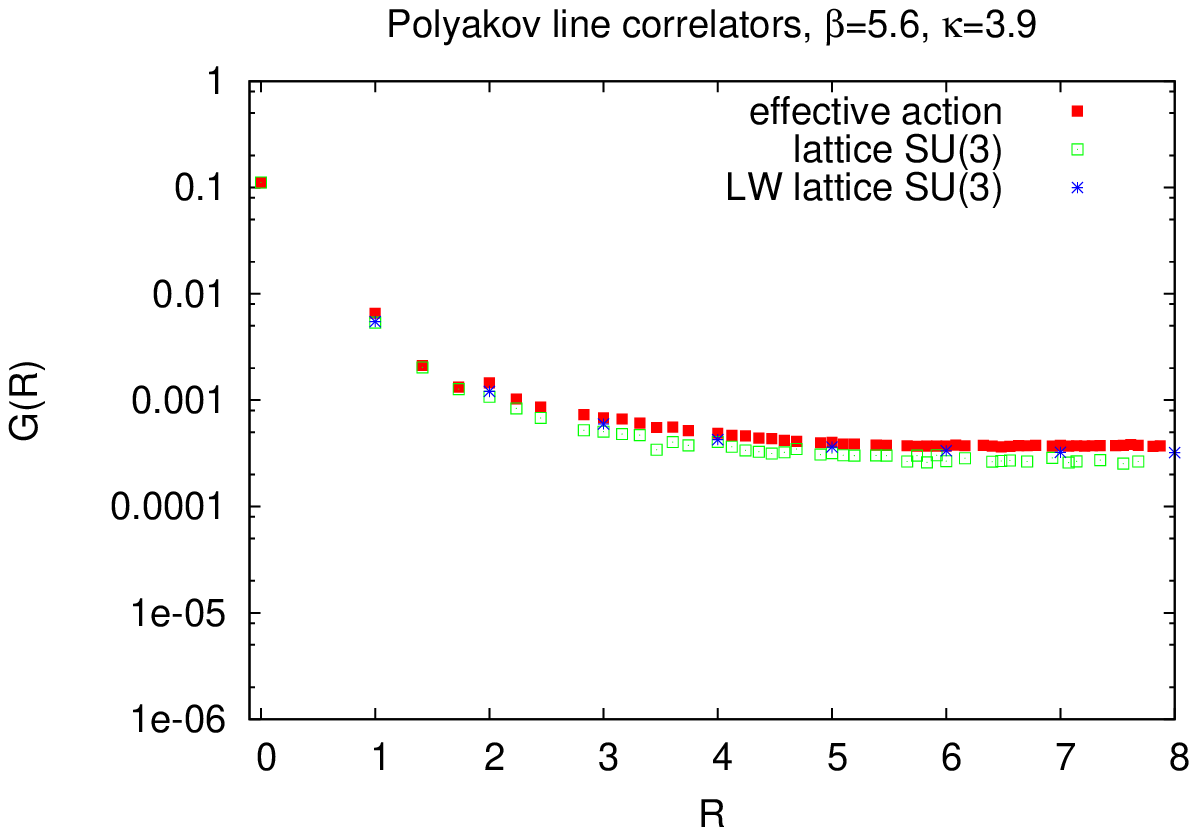}}}
\caption{The Polyakov line correlators for the gauge-Higgs theory at $\b=5.6$ and $\k=3.9$,  corresponding to the
lightest matter field in our set of $\k$ values, computed from numerical simulation of the effective PLA $S_P$, and from simulation of the underlying lattice SU(3) gauge theory.  On-axis data points denoted ``LW lattice SU(3)'' are derived from the
underlying theory with L\"uscher-Weisz noise reduction.}
\label{corr56k39}
\end{figure}

   The comparison of off-axis Polyakov line correlators at $\b=5.6, \kappa=3.9$ computed for $S_P$ and for the underlying lattice gauge-Higgs theory is shown in Fig.\ \ref{corr56k39}.  On-axis data points derived from the underlying theory using L\"uscher-Weisz noise reduction \cite{Luscher:2001up} are also displayed in this figure.

\subsection{$\mathbf \k=3.8, 3.6$}

     As $\k$ is reduced, the effective theory should approach the pure gauge result discussed in Section \ref{pure}.
Even a small reduction away from the crossover, from $\k=3.9$ to $\k=3.8$ has a large effect on the Polyakov line
correlator, as we see in Fig.\ \ref{pot56kappa}.  

\begin{figure}[htb]
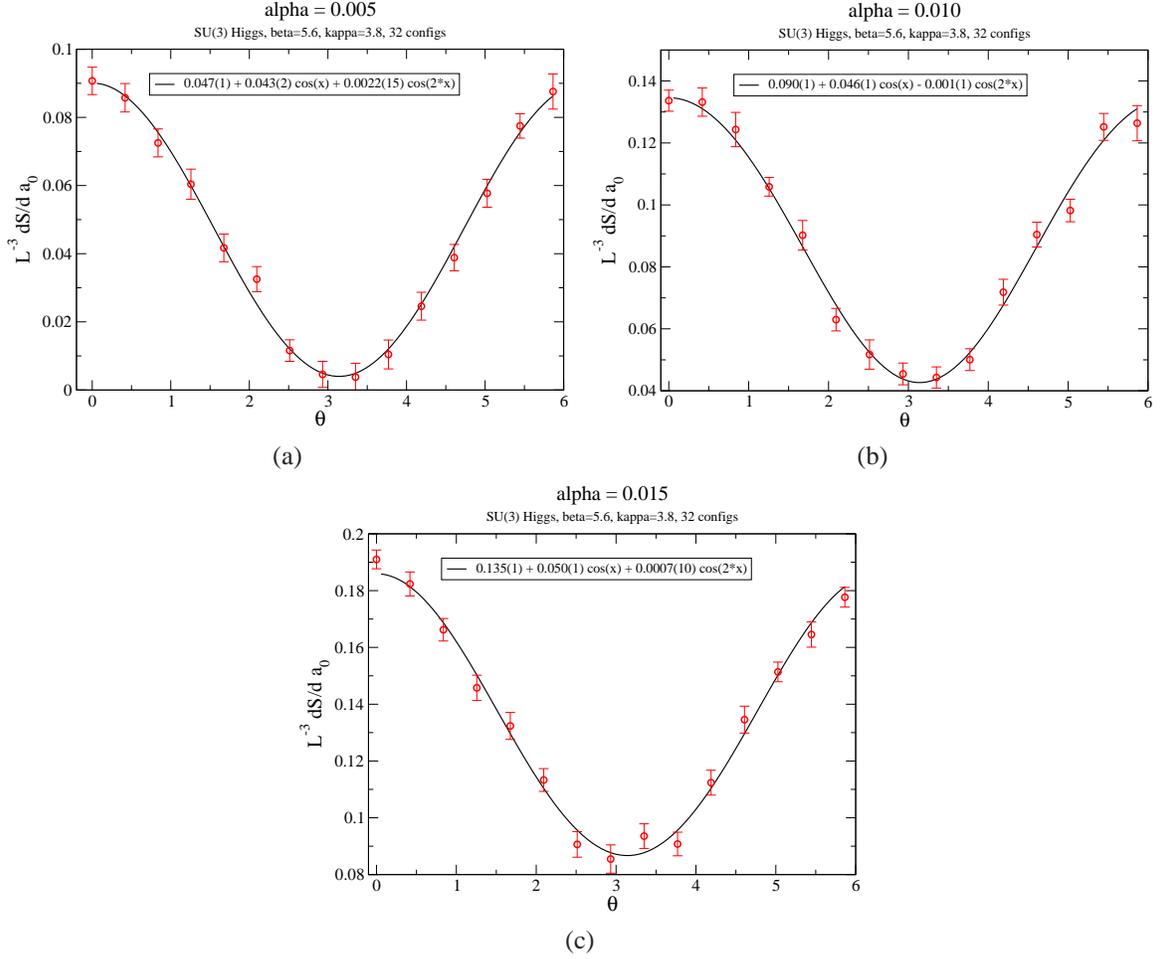

\subfigure[]  
{   
 \label{}
 \includegraphics[scale=0.3]{mu05k38.eps}
}
\subfigure[]  
{   
 \label{}
 \includegraphics[scale=0.3]{mu10k38.eps}
}
\subfigure[]  
{   
 \label{}
 \includegraphics[scale=0.3]{mu15k38.eps}
}
\caption{Same as Fig.\ \ref{muk39}, but this time at $\k=3.8$. (a) $\a$=0.005; (b) $\a$=0.010; (c) $\a$=0.015.}
\label{muk38}
\end{figure}

\begin{figure}[htb]
\centerline{\scalebox{0.9}{\includegraphics{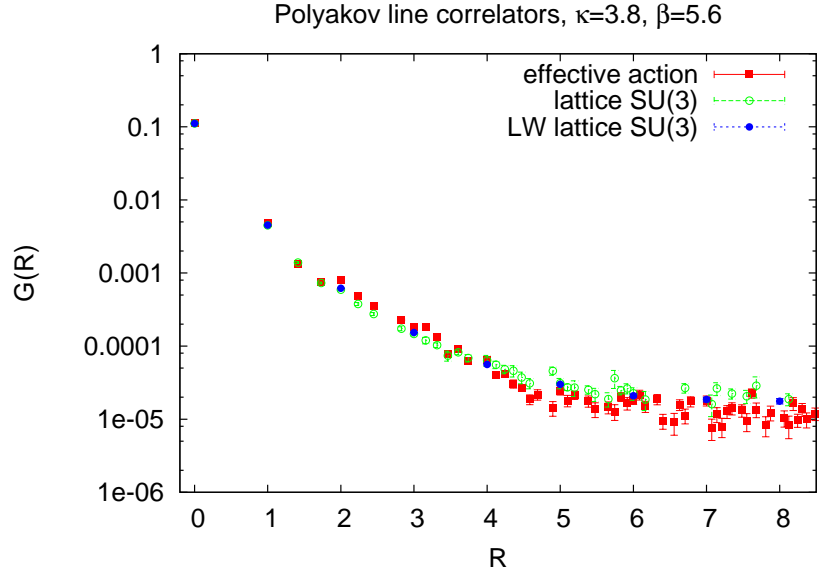}}}
\caption{The Polyakov line correlators for the gauge-Higgs theory at $\b=5.6$ and $\k=3.8$,  computed from numerical simulation of the effective PLA $S_P$, and from simulation of the underlying lattice SU(3) gauge theory. In the latter case
we show off-axis points computed by standard methods, together with on-axis points using L\"uscher-Weisz noise reduction.}
\label{corr56k38}
\end{figure}

    The effective action at $\k=3.8$ is determined by the same means as at the larger $\k=3.9$ value.   The main difference
is that the center-symmetry breaking terms proportional to $e^{\pm 2i\th}$ are consistent with zero, within error bars.  We
only show the results, in Fig.\  \ref{muk38}, for the zero-mode derivative, which can be compared to Fig.\ \ref{muk39} above.
Note that the coefficient of the $\cos(2\th)$ term is essentially consistent with zero.  It is unlikely that this and higher terms
in fugacity are {\it exactly} zero, but they are too small to be detected with our current statistics.  The Polyakov line
correlator comparison at $\k=3.8$ is displayed in Fig.\ \ref{corr56k38}.

\begin{figure}[htb]
\centerline{\scalebox{0.9}{\includegraphics{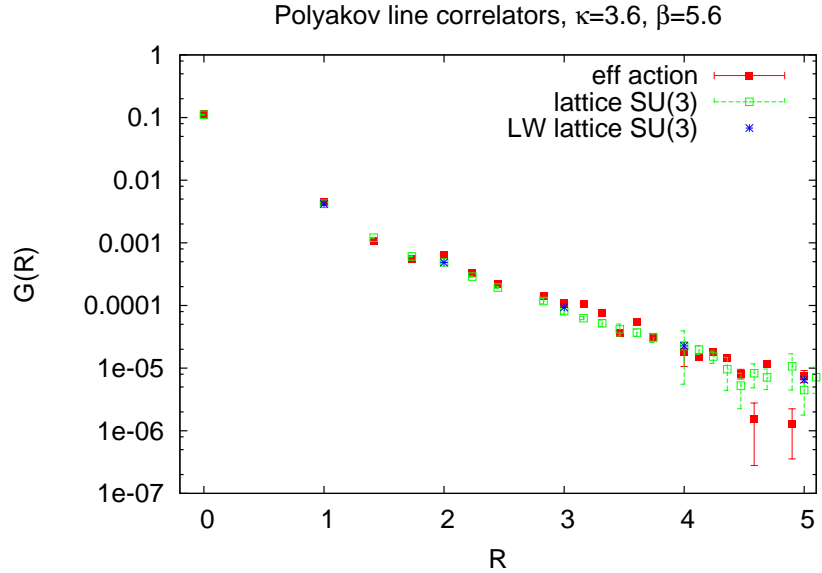}}}
\caption{The Polyakov line correlators for the gauge-Higgs theory at $\b=5.6$ and $\k=3.6$,  corresponding to the heaviest scalar in our set of $\k$ values, computed from numerical simulation of the effective PLA $S_P$, and from simulation of the underlying lattice SU(3) gauge theory.}
\label{corr56k36}
\end{figure}

We have also carried out our procedure for $\k=3.6$, and the corresponding correlator comparison is shown in Fig.\ 
\ref{corr56k36}.  In this case the mass of the matter field is so large that the results are not far from the pure-gauge result at 
$\b=5.6$.  The parameters which determine the effective action $S_P$ at $\b=5.6$ and $\k=3.6,3.8,3.9$ are shown in Table 
\ref{tab2}.

\clearpage

\begin{table}[t!]
\begin{center}
\begin{tabular}{|c|c|c|c|c|c|c|c|c|} \hline
         $ \k $ &  $c_1$ & $c_2$ & $k_0$ &  $b_1$ & $b_2$ & $r_{max}$ & $d_1$ & $d_2$ \\
\hline
        3.6 &  8.53(6) &  0.99(4) & 1.68 & 6.68(14) & 0.71(2) & $\sqrt{39}$ & 0.0062(7) & $< 0.001$ \\
        3.8 &  9.77(8) & 1.18(2)  & 1.63 & 6.77(17) & 0.72(2)   & $\sqrt{41}$ & 0.0195(4) & $< 0.001$  \\ 
        3.9 &  12.55(13) & 1.69(4) & 1.36 & 8.16(17) & 0.89(2) & no cutoff  & 0.0585(8)  & 0.0115(2)  \\          
\hline
\end{tabular}
\caption{Parameters defining the effective Polyakov line action $S_P$ for SU(3) gauge-Higgs theory at ${\b=5.6}$ 
and $\k=3.6,3.8,3.9$ on a $16^3 \times 6$ lattice.} 
\label{tab2}
\end{center}
\end{table}

\section{\label{mf}Mean field approach to solving the effective action}

   In this section we solve the effective theory $S_P$ in \rf{SP_for_ghiggs}, derived for the gauge-Higgs action \rf{ghiggs}
at finite chemical potential, and also for the effective theory derived for heavy quarks at large chemical potential.
In both cases the effective action $S_P$ still has a sign problem.  As noted in the Introduction, the sign problem in the effective model can be attacked by a variety of methods \cite{Mercado:2012ue,Aarts:2011zn,Fromm:2011qi,Greensite:2012xv}, which have been successfully applied to the SU(3) spin model \rf{spin}.  Here we will implement the mean field approach, following closely the treatment in \cite{Greensite:2012xv}, and postponing the treatment by other procedures to later work.  The mean field method is, of course, an approximation, but it is worth noting that the approximation typically improves the more spins are coupled, in the action, to any given spin. For an action such as $S_P$, in which not only nearest neighbor spins, but spins separated by any distance $\le r_{max}$ are coupled together, it is possible that the mean field treatment provides a better approximation than
one might otherwise expect in $D=3$ dimensions. 

\subsection{The gauge-Higgs model}  

   The starting point is the effective bilinear action \rf{SP_for_ghiggs}, where $K(\vx-\vy)$ is determined from the parameters in 
Table \ref{tab2}. While $Q(\vx-\vy,\mu)$ is consistent with zero, at the level of our present statistics, we will carry it along
just to show how it is included in the mean field approach.  Reintroducing the holonomies via the definition \rf{P}, the bilinear
action has the form
\bea
S_P &=& \sum_{xy} \tr U_\vx \tr U_\vy^\dg \on K(\vx-\vy) + \sum_\vx \Bigl\{ \ot(d_1 e^{\m/T} - d_2 e^{-2\m/T}) \tr U_\vx 
  + \ot (d_1 e^{-\m/T} - d_2 e^{2\m/T}) \tr U^\dg_\vx \Bigr\}
\non \\
& & + \sum_{xy} \Bigl(\tr U_\vx \tr U_\vy \on Q(\vx-\vy,\mu) +  \tr U^\dg_\vx \tr U^\dg_\vy \on Q(\vx-\vy;-\m) \Bigr) \ .
\eea
Introducing a notation for the double sum over sites $\vx,\vy$ that excludes $\vx=\vy$
\beq
    \sum_{(\vx,\vy)} \equiv \sum_{\vx} \sum_{\vy \ne \vx}
\eeq
we have
\bea
S_P &=& \sum_{(\vx,\vy)} \tr U_\vx \tr U_\vy^\dg \on K(\vx-\vy) + \sum_{\vx} \tr U_\vx \tr U_\vx^\dg \on K(0)
\non \\
& & +  \sum_\vx \Bigl\{ \ot(d_1 e^{\m/T} - d_2 e^{-2\m/T}) \tr U_\vx 
  + \ot (d_1 e^{-\m/T} - d_2 e^{2\m/T}) \tr U^\dg_\vx \Bigr\}
\non \\
& & + \sum_{(\vx,\vy)} (\tr U_\vx \tr U_\vy \on Q(\vx-\vy,\mu) +  \tr U^\dg_\vx \tr U^\dg_\vy \on Q(\vx-\vy;-\m) )
\non \\
& & + \sum_{\vx} (\tr U_\vx \tr U_\vx \on Q(0,\mu) +  \tr U^\dg_\vx \tr U^\dg_\vx \on Q(0;-\m) ) \ .
\eea
Let us focus on the two semi-local terms
\bea
T_1  &=& \sum_{(\vx,\vy)} \tr U_\vx \tr U_\vy^\dg \on K(\vx-\vy) \ ,
\non \\
T_2 &=& \sum_{(\vx,\vy)} (\tr U_\vx \tr U_\vy \on Q(\vx-\vy,\mu) +  \tr U^\dg_\vx \tr U^\dg_\vy \on Q(\vx-\vy;-\m) ) \ ,
\eea
and write
\beq
   \tr U_\vx = (\tr U_\vx - u) + u  ~~~,~~~ \tr U^\dg_\vx = (\tr U^\dg_\vx - v) + v \ .
\eeq
Then
\bea
T_1 &=& \sum_{(\vx \vy)} \Bigl\{ u \tr U_\vy^\dg + v \tr U_\vx - uv \Big\} \left(\on K(\vx-\vy) \right) + E_1
\non \\
&=& J_0 \sum_\vx (v \tr U_\vx + u \tr U_\vx^\dg) - uvJ_0V + E_1 \ ,
\eea
where we have defined
\bea
     E_1 &=& \sum_{(\vx \vy)} (\tr U_x-u)(\tr U_\vy^\dg - v) \on K(\vx-\vy) \ ,
\non \\
     J_0 &=& \on \sum_{\vx \ne 0} K(\vx) \ .
\eea
Likewise
\bea
T_2 &=& 2 \sum_\vx (u \tr U_\vx J_2(\m) + v \tr U^\dg_\vx J_2(-\m) - (u^2 J_2(\m) + v^2 J_2(-\m))V + E_2 \ ,
\eea
where
\bea
E_2 &=&  \sum_{(\vx,\vy)} \Bigl\{ (\tr U_\vx-u) (\tr U_\vy-u) \on Q(\vx-\vy,\mu) +  (U^\dg_\vx-v) (\tr U^\dg_\vy-v) 
\on Q(\vx-\vy;-\m) \Bigr\} \ ,
\non \\
J_2(\m) &=& \on \sum_{\vx \ne 0} Q(\vx,\m) ~~,~~ J_2(-\m) = \on \sum_{\vx\ne 0} Q(\vx,-\m) \ .
\eea
Putting it all together,
\bea
S_P &=& \sum_\vx \tr U_\vx \Bigl\{ J_0 v + \ot(d_1 e^{\m/T} - d_2 e^{-2\m/T} + 2 J_2(\m) u \Bigr\}
\non \\
& & + \sum_\vx \tr U^\dg_\vx \Bigl\{ J_0 u + \ot(d_1 e^{-\m/T} - d_2 e^{2\m/T} + 2 J_2(-\m) v \Bigr\}
\non \\
& & - u v J_0 V - (u^2 J_2(\m) + v^2 J_2(-\m) )V +  \sum_\vx \tr U_x \tr U^\dg_x \on K(0)
\non \\
& & + \on \sum_\vx \Bigl\{ \tr U_\vx \tr U_\vx Q(0,\m) + \tr U^\dg_\vx \tr U^\dg_\vx Q(0,-\m) \Bigr\} + E_1 + E_2 \ .
\eea

   The mean field approximation amounts to dropping $E_1,E_2$, and then choosing the constants $u,v$ such that
the free energy of the resulting theory is minimized.  The justification is that $E_1,E_2$ depend only on the differences
$\tr U_\vx - u$ and $\tr U^\dg_\vx - v$,  and the choice of $u,v$ minimizing the free energy sets the expectation value of
these differences to zero.  The approximation can be improved by treating $E_1,E_2$ as 
small corrections to the leading mean field result, as carried out for the SU(3) spin model in \cite{Greensite:2012xv}, but 
for now we will just work in the leading approximation, neglecting $E_1,E_2$.

   Let us define
\bea
A(\m) &=& J_0 v + \ot(d_1 e^{\m/T} - d_2 e^{-2\m/T}) + 2 J_2(\m) u \ ,
\non \\
B(\m) &=& J_0 u + \ot(d_1 e^{-\m/T} - d_2 e^{2\m/T}) + 2 J_2(-\m) v \ ,
\non \\
a_0 &=& \on K(0) ~~~,~~~ a_2(\m) = \on Q(0,\m) ~~~,~~~ a_2(-\m) = \on Q(0,-\m) \ .
\label{AB}
\eea
The partition function of the effective model, in the mean field approximation, is then
\bea
Z_{mf} &=& \exp\Bigl[-uvJ_0 V - (u^2 J_2(\m) + v^2 J_2(-\m))V \Bigr]
\non \\
&\times& \left\{ \exp\left[a_0 {\pa^2 \over \pa A \pa B} + a_2(\m){\pa^2 \over \pa A^2} + a_2(-\m){\pa^2 \over \pa B^2}\right] 
      \int DU e^{A \tr U + B \tr U^\dg} \right\}^V \ .
\eea
We introduce the rescalings
\beq
    u = u' e^{-\m/T} ~~,~~ v = v' e^{\m/T} ~~,~~ A = A' e^{\m/T} ~~,~~ B = B' e^{-\m/T} \ ,
\label{rescale}
\eeq
and follow the steps in ref.\ \cite{Greensite:2012xv}, which will not be reproduced here.
The upshot is that if we denote $Z_{mf} = \exp[-f_{mf} V/T]$, where $V$ is the lattice volume in $D=3$ dimensions, then
\bea
f_{mf}/T = u' v' J_0 + u'^2 e^{-2\m/T} J_2(\m) + v'^2 e^{2\m/T} J_2(-\m) - \log F[A',B'] \ ,
\eea
where\footnote{In practice $F[A',B']$ is evaluated by expanding the exponential containing differential operators in a Taylor series, and truncating the series.  In this particular gauge-Higgs example, $a_0$ is very small compared to $J_0$, and the expansion to first order makes hardly any difference to the result at zeroth order.  The sum over $s$ is also truncated to
$|s| \le s_{max}$, and we have checked the increasing the cutoff beyond $s_{max}=3$ makes no difference to the result.}
\bea
F[A',B'] &=& \exp\left[a_0 {\pa^2 \over \pa A' \pa B'} + a_2(\m)e^{-2\m/T}{\pa^2 \over \pa A'^2} 
                      + a_2(-\m)e^{2\m/T}{\pa^2 \over \pa B'^2}\right] 
\non \\
& & \times \sum_{s=-\infty}^{\infty} e^{3 \mu s} \det\Bigl[D^{-s}_{ij} I_0[2\sqrt{A' B'}] \Big] \ ,
\eea
and $D^{-s}_{ij}$ is the $i,j$-th component of a matrix of differential operators
\bea
D^s_{ij} &=& \left\{ \begin{array}{cl}
                         D_{i,j+s} & s \ge 0 \cr
                         D_{i+|s|,j} & s < 0 \end{array} \right. \ ,
\non \\
D_{ij} &=& \left\{ \begin{array}{cl}
                         \left({\pa \over \pa B'} \right)^{i-j} & i \ge j \cr 
                        \left({\pa \over \pa A'} \right)^{j-i} & i < j \end{array} \right. \ .
\eea
Since \rf{AB} can be inverted to give $u',v'$ in terms of $A',B'$, we find the minimum of $f_{mf}$ by solving the
stationarity conditions
\bea
\left\{ {\pa u' \over \pa A'} v' + u' {\pa v' \over \pa A'}\right) J_0 
 + 2\left( u' {\pa u' \over \pa A'} \right) e^{-2\m/T} J_2(\m)  + 2v'\left( {\pa v' \over \pa A'}\right)e^{2\m/T} J_2(-\m)
   - {1\over F}{\pa F \over \pa A'} &=& 0 \ ,
\non \\
\left\{ {\pa u' \over \pa B'} v' + u' {\pa v' \over \pa B'}\right) J_0 
 + 2\left( u' {\pa u' \over \pa B'} \right) e^{-2\m/T} J_2(\m)  + 2v'\left( {\pa v' \over \pa B'}\right)e^{2\m/T} J_2(-\m)
   - {1\over F}{\pa F \over \pa B'} &=& 0 \ ,
\non \\
\eea
numerically. 

\begin{figure}[htb]
\subfigure[]  
{   
 \label{uv}
 \includegraphics[scale=0.6]{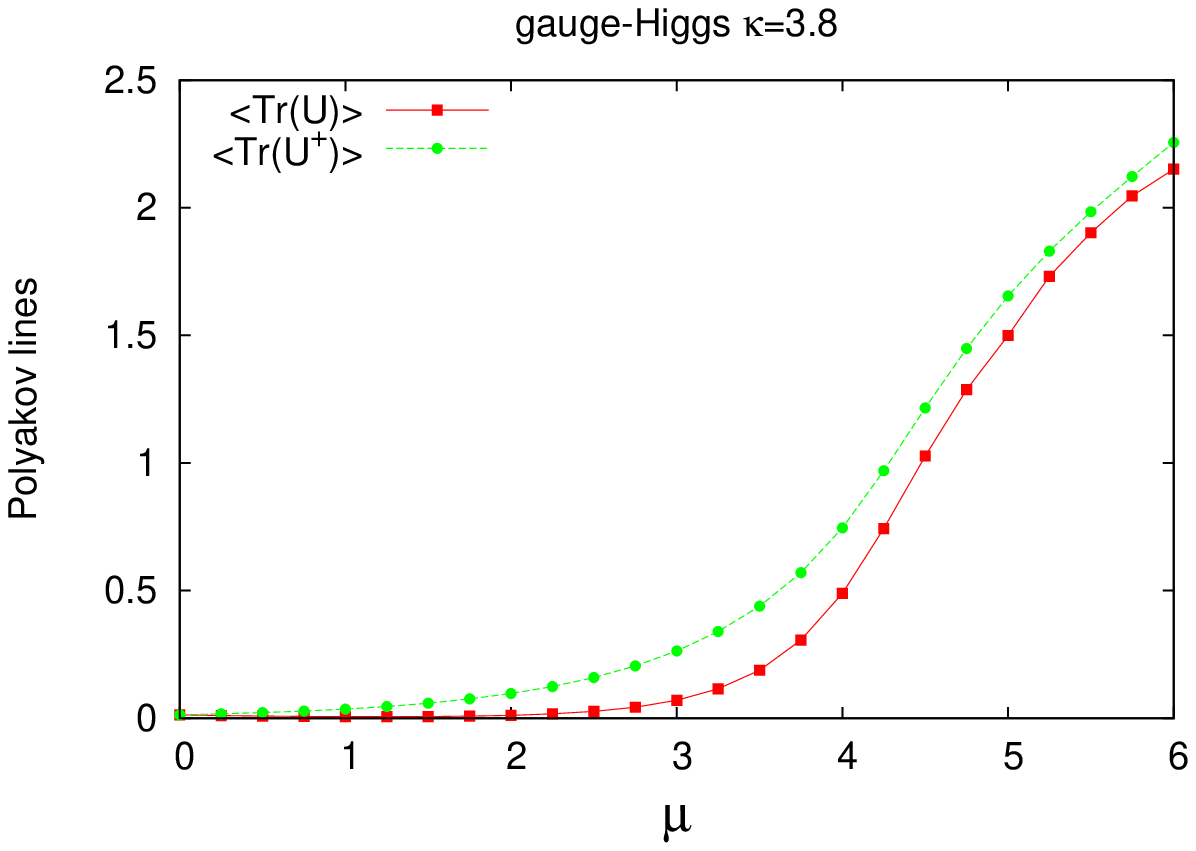}
}
\subfigure[]  
{   
 \label{nden}
 \includegraphics[scale=0.6]{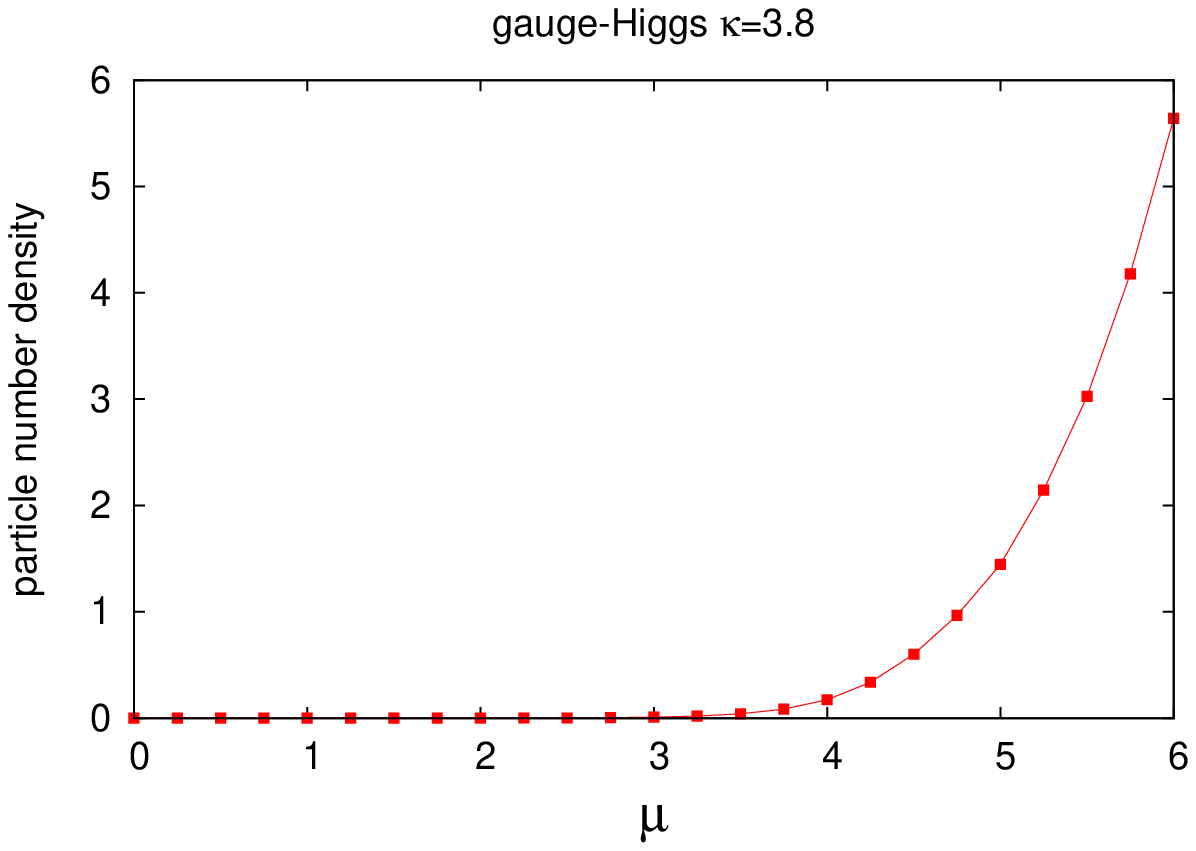}
}
\caption{Mean field solution of the effective Polyakov line action $S_P$ corresponding to a gauge-Higgs theory at $\b=5.6$, 
$\k=3.8$, at finite values of the chemical potential.  (a) the expectation value of Polyakov lines 
$\langle \tr U \rangle$ and $\langle \tr U^\dg \rangle$ vs.\ $\m/T$; (b) particle number density vs.\ $\m/T$.}
\label{uvn}
\end{figure}

   For the present we are ignoring the $Q(\vx-\vy)$ kernel, which is certainly negligible at small to moderate $\m$.
In this case one can show that
\bea
u' &=&  J_0  {\pa \over \pa A'} (u' v') \ ,
\non \\
v' &=&  J_0  {\pa \over \pa B'} (u' v') \ ,
\eea
and the stationarity conditions simplify to
\bea
u' - {1\over F} {\pa F \over \pa A'} &=& 0 \ ,
\non \\
v' - {1\over F} {\pa F \over \pa B'} &=& 0 \ .
\eea
But we also have, in the mean field approximation, that \cite{Greensite:2012xv}
\bea
\langle \tr U_x \rangle =  {1\over F}{\pa F \over \pa A}  ~~,~~  \langle \tr U^\dg_x \rangle =  {1\over F}{\pa F \over \pa B} \ ,
\eea
which, together with the stationarity conditions, imply the self-consistency conditions
\bea
u = \langle \tr U_x \rangle ~~~,~~~ v = \langle \tr U^\dg_x \rangle \ .
\eea
For phase structure the relevant observables are $u,v$ and the scalar ``quark'' number density
\bea
            n &=& - {d f_{mf} \over d \m} = -T\left({\pa \over \pa \m} + {\pa A' \over \pa \m}{\pa \over \pa A'} 
            +  {\pa B' \over \pa \m}{\pa \over \pa B'} \right) f_{mf} 
\non \\
 &=&  {1\over F} {\pa F \over \pa \mu/T} 
\eea
where $f_{mf}$ is evaluated at the stationary point, so that derivatives of $f_{mf}$ wrt $A',B'$ vanish.
These observables are plotted as a function of $\m/T$ in Fig.\ \ref{uvn} for the case of $\b=5.6, \k=3.8$.  We see no evidence of a phase transition.

   The case of $\b=5.6, \k=3.9$ is more problematic.  In this case, the mean field solution yields a negative number density
at finite $\m$, which we consider to be an unphysical result.  The error may lie in the mean field method itself, but more
likely it is due to the neglect of center symmetry-breaking terms which are bilinear in the Polyakov lines.  Although such terms appear to be unimportant at $\m=0$, we can see from our data (e.g.\ Fig.\ \ref{Q100}) that they exist, and presumably become relevant at finite $\m$.  We will return to this example, and a comparison of mean field and complex Langevin techniques, in a subsequent article \cite{Me}.

\subsection{The heavy quark model}

   Let $\zeta$ represent the hopping parameter for Wilson fermions, or $1/2m$ for staggered fermions, and $h=\zeta^{N_t}$.
In the limit that $\zeta \ra 0$ and $e^\mu \ra \infty$ in such a way that $\zeta e^\mu $ is finite, the lattice action simplifies drastically \cite{Bender:1992gn,*Blum:1995cb,*Engels:1999tz,*DePietri:2007ak}.  In temporal gauge,
\bea
\exp[S_L] = \prod_\vx \det\Bigl[1+h e^{\m/T} U_0(\vx,0)\Bigr]^p \det\Bigl[1+h e^{-\m/T} U^\dg(\vx,0) \Bigr]^p \exp[S_{plaq}] \ ,
\eea
where $p=1$ for four-flavor staggered fermions, and $p=2N_f$ for Wilson fermions ($N_f$ is the number of flavors), and where
the determinant refers to color indices since the Dirac indices have already been accounted for.  Since the determinants only involve the Polyakov loop holonomies, the effective PLA is derived trivially once one has derived the $S_P^{pg}$ for the pure gauge theory defined by the plaquette action $S_{plaq}$:
\bea
\exp[S_P] = \prod_\vx \det\Bigl[1+h e^{\m/T} U_\vx \Bigr]^p \det\Bigl[1+h e^{-\m/T} U^\dg_\vx \Bigr]^p   \exp[S_P^{pg}] \ .
\eea
The determinants can be expressed entirely in terms of Polyakov line operators, using the identities
\bea
\det\Bigl[1+h e^{\m/T} U_\vx \Bigr] &=& 1 + he^{\m/T} \tr[U_\vx] + h^2 e^{2\m/T} \tr[U_\vx^\dg] + h^3 e^{3\m/T} \ ,
\non \\
\det\Bigl[1+h e^{-\m/T} U^\dg_\vx \Bigr] &=& 1 + he^{-\m/T} \tr[U_\vx^\dg] + h^2 e^{-2\m/T} \tr[U_\vx] + h^3 e^{-3\m/T} \ .
\eea
This leads us to the mean field expression\footnote{A term $a_0 {\pa^2 \over \pa A \pa B}$ in the leading exponential containing 
$J_0 uv$ is neglected, since $a_0$ is two orders of magnitude smaller than $J_0$.}
\bea
      Z_{mf} &=& \left\{ e^{-J_0 uv}\int dU \left(1 + he^{\m/T} \tr[U] + h^2 e^{2\m/T} \tr[U^\dg] + h^3 e^{3\m/T}\right)^p \right.
\non \\
 & & \times \left.  \left(1 + he^{-\m/T} \tr[U^\dg] + h^2 e^{-2\m/T} \tr[U] + 
 h^3 e^{-3\m/T}\right)^p  \exp[A\tr U + B\tr U^\dg] \right\}^V
 \non \\
 &=&  \left\{ e^{-J_0 uv}\left(1 + he^{\m/T} {\pa \over \pa A} + h^2 e^{2\m/T} {\pa \over \pa B} + h^3 e^{3\m/T}\right)^p \right.
\non \\
 & & \times \left.  \left(1 + he^{-\m/T} {\pa \over \pa B} + h^2 e^{-2\m/T} {\pa \over \pa A}+ 
 h^3 e^{-3\m/T}\right)^p  \int dU \exp[A\tr U + B\tr U^\dg] \right\}^V
\non \\
&=&  \left\{ e^{-J_0 u'v' }\left(a_1 + a_2 e^{-\m/T} {\pa \over \pa A'} + a_3 e^{\m/T} {\pa \over \pa B'}    
 +a_4 e^{-2\m/T} {\pa^2 \over \pa A'^2} \right. \right.
 \non \\
& & \qquad \left. \left. + a_5 e^{2\m/T} {\pa^2 \over \pa B'^2} + a_6 {\pa^2 \over \pa A' \pa B'} \right)^p \right.
 \left. \times \sum_{s=-\infty}^{\infty} e^{3 \mu s} \det\Bigl[D^{-s}_{ij} I_0[2\sqrt{A' B'}] \Big] \right\}^V \ ,
 \eea
where
\bea
a_1 &=& 1 + h^3 (e^{3\mu/T}+e^{-3\mu/T}) + h^6
\non \\
a_2 &=&  (h+h^5)e^{\m/T} + (h^2+h^4)e^{-2\m/T} ~~~,~~~ a_3 =  (h+h^5)e^{-\m/T} + (h^2+h^4)e^{2\m/T} 
\non \\
a_4 &=& h^3 e^{-\m/T} ~~~,~~~ a_5 = h^3 e^{\m/T} ~~~,~~~ a_6 = h^2 + h^3 \ ,
\eea
and in this case $A = J_0 v, ~ B = J_0 u$, with rescalings as in \rf{rescale}.   Defining 
\bea
G(A',B') &=& \left(a_1 + a_2 e^{-\m/T} {\pa \over \pa A'} + a_3 e^{\m/T} {\pa \over \pa B'} +   
 + a_4 e^{-2\m/T} {\pa^2 \over \pa A'^2} \right.
\non \\
& & \left. + a_5 e^{2\m/T} {\pa^2 \over \pa B'^2}  + a_6 {\pa^2 \over \pa A' \pa B'}\right)^p
 \sum_{s=-\infty}^{\infty} e^{3 \mu s} \det\Bigl[D^{-s}_{ij} I_0[2\sqrt{A' B'}] \Big] \ ,
\eea
then the mean field self-consistency conditions $u=\langle \tr U_\vx \rangle, v=\langle \tr U^\dg_\vx \rangle$, equivalent to a stationarity condition on the mean field free energy, are
\bea
{B' \over J_0} - {1\over G}{\pa G \over \pa A'}=0  ~~~~~\text{and}~~~~~ {A' \over J_0} - {1\over G}{\pa G \over \pa B'}=0 \ ,
\eea
which can be solved numerically.

\begin{figure}[htb]
\subfigure[]  
{   
 \label{uv_S}
 \includegraphics[scale=0.6]{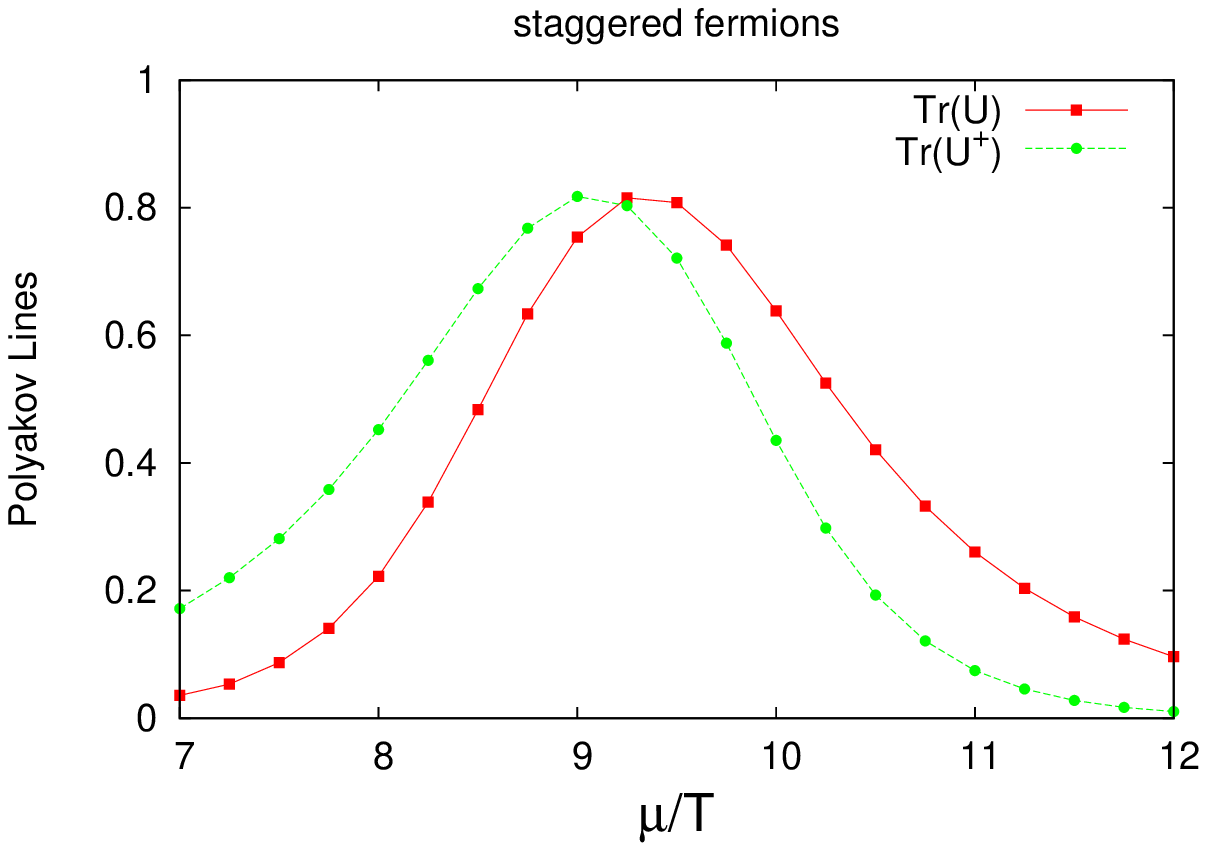}
}
\subfigure[]  
{   
 \label{nden_S}
 \includegraphics[scale=0.6]{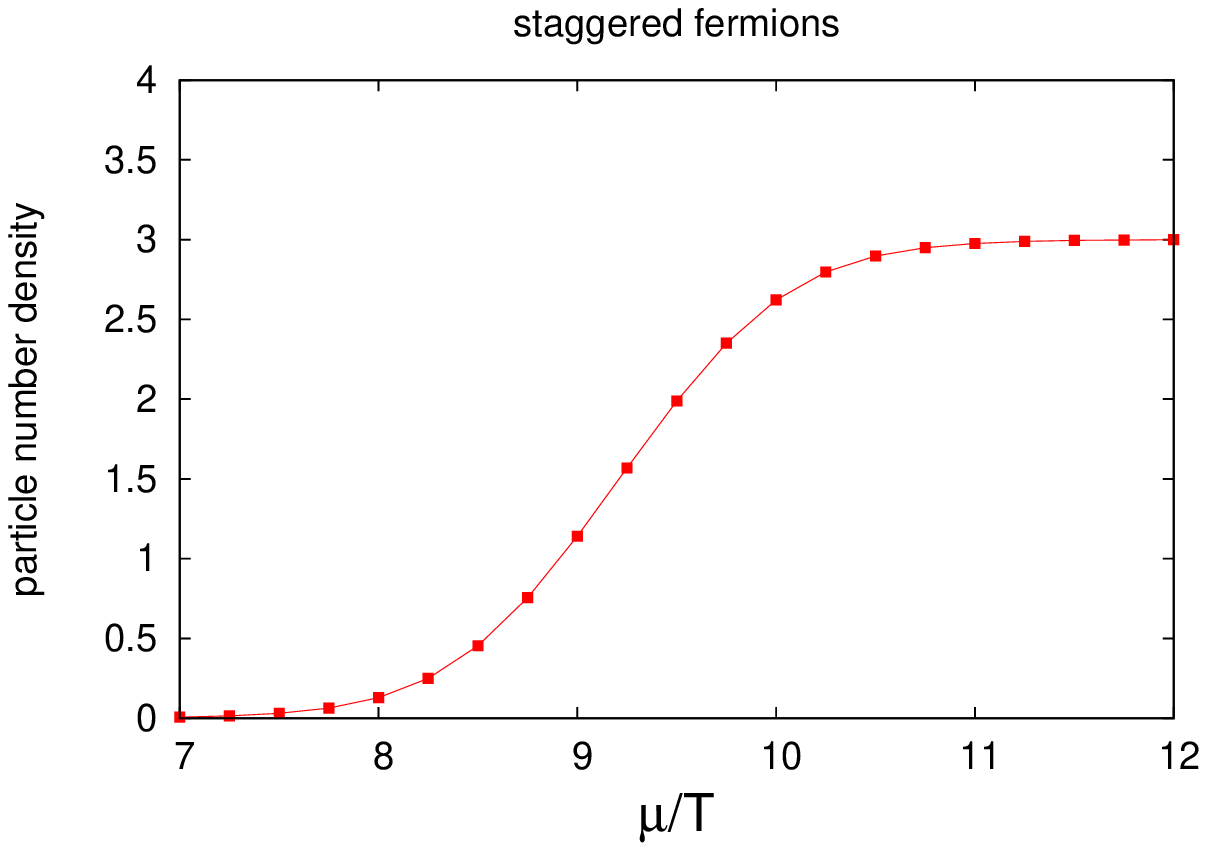}
}
\caption{Mean field solution of the effective Polyakov line action $S_P$ corresponding to a gauge theory on a
$16^3 \times 6$ lattice at $\b=5.6$, with heavy staggered fermions of mass $m=2.32$ in lattice units.  (a) the expectation value of Polyakov lines  $\langle \tr U \rangle$ and $\langle \tr U^\dg \rangle$ vs.\ $\m/T$; (b) particle number density vs.\ $\m/T$.}
\label{uvnS}
\end{figure}

    As an example, we have solved the heavy quark model for staggered quarks ($p=1$, four flavors) at $\b=5.6, N_t=6$ and
$h=10^{-4}$, which corresponds to a mass $m=2.32$ in inverse lattice spacing.  The result is shown in Fig.\ \ref{uvnS}.
Note that the number density saturates  for large $\m/T$ at $n=3$ particles/lattice site, as is appropriate for staggered quarks with three colors.

\section{\label{conclude}Conclusions}

    We have tested the relative weights method for extracting the effective Polyakov line action from both pure SU(3) lattice gauge theory and in an SU(3) gauge-Higgs theory in the ``confinement-like'' phase.  In the latter case we have shown how to compute the effective action also in the case of finite chemical potential.  In all cases studied so far there is excellent agreement between Polyakov line correlators computed in the effective action and in the underlying gauge theory at zero chemical potential.  Mean field methods have been employed to determine the expectation value of observables in the effective action, corresponding to the gauge-Higgs theory \rf{ghiggs} at $\b=5.6, \k=3.8$, and to a gauge theory with massive quarks, 
as a function of chemical potential.
    
    So far we have computed the effective action up to terms bilinear in the Polyakov lines, at zeroth order in fugacity, and terms linear in the Polyakov lines, up to second order in the fugacity.  It is straighforward, i.e.\ only a matter of increased statistics, to extract also the fugacity dependent bilinear terms.  We believe that the method can be extended to derive terms involving products of three or four Polyakov lines by fitting the path derivatives \rf{O} to polynomials in $\a$, and by computing second derivatives of $S_P$ with regard to momentum modes.  It is important to determine at least the magnitude of $\m$-dependent terms which are neglected at $\m=0$, as compared to terms which are kept, because this will give us an estimate of how far out we can go in $\m$ before the neglected terms become important.  This problem is currently under investigation.

    The next steps in our program are as follows:  First, since any effective Polyakov line action at finite $\m$ has a sign problem,
it is essential to assess the reliability of mean field theory in this context, or to find another technique, such as complex Langevin \cite{Aarts:2011zn} or the density of states method \cite{Langfeld:2014nta,*Langfeld:2012ah}, to deal with the problem.  One thing we can do along these lines is to compare mean field and complex Langevin solutions of the effective actions we have derived so far.  This work is well underway, and the results will be reported shortly \cite{Me}.  We would then like to extract terms in the effective action for gauge-Higgs theory, such as bilinear terms to second order in fugacity, which have been neglected so far.  The final step is to replace the scalar field with fermion fields, and solve for the effective Polyakov line action.  The application of our method to the
case of gauge fields coupled to fermions was already outlined in the appendix of ref.\ \cite{Greensite:2013yd}.  We have now seen that the introduction of an imaginary chemical potential is essential, and this technique should supplement the approach in \cite{Greensite:2013yd}.   One can then vary parameters, and search for phase transitions.  Of course the ultimate goal, if it proves feasible by these methods, is work out at least some of the phase diagram in the $\m-T$ plane for SU(3) lattice gauge fields coupled to light dynamical quarks; i.e.\ QCD.  The work reported in this article is intended as one of the necessary steps in that direction.

\acknowledgments{J.G.'s research is supported in part by the
U.S.\ Department of Energy under Grant No.\ DE-FG03-92ER40711.  K.L.'s research is supported by STFC under the DiRAC framework. We are grateful for support from the HPCC Plymouth, where the numerical computations have been carried out.}     
   
\bibliography{pline}
 
\end{document}